\documentclass[useAMS,usenatbib]{mn2e}
\usepackage[english]{babel}

\usepackage{fancyhdr}
\usepackage{amsfonts}
\usepackage{amsmath}
\usepackage{amssymb}
\usepackage{multicol}
\usepackage{layout}
\usepackage{graphicx}

\usepackage{times}
\usepackage{natbib}

\newif\ifAMStwofonts
\AMStwofontstrue

\graphicspath{{./fig/}}

\title[Structure growth with oscillating dark energy]{Structure formation in cosmologies with oscillating dark energy}

\author[F. Pace et al.]
       {F. Pace$^{1,2}$\thanks{E-mail: Francesco.Pace@port.ac.uk},
        C. Fedeli$^{3}$, L. Moscardini$^{4,5,6}$ and M. Bartelmann$^{2}$\\
        $^{1}$ Institute of Cosmology and Gravitation, University of Portsmouth, Dennis Sciama Building, Portsmouth, PO1 3FX, U.K.\\
        $^{2}$ Zentrum f\"ur Astronomie der Universit\"at Heidelberg, Institut f\"ur Theoretische Astrophysik, Albert-Ueberle-Str. 2, D-69120 Heidelberg, Germany \\
	$^{3}$ Department of Astronomy, University of Florida, 211 Bryant Space Science Center, Gainesville, FL 32611-2055, USA\\
	$^{4}$ Dipartimento di Astronomia, Universit\`a di Bologna, Via Ranzani 1, I-40127 Bologna, Italy \\
        $^{5}$ INFN, Sezione di Bologna, Viale Berti Pichat 6/2, I-40127 Bologna, Italy \\
	$^{6}$ INAF, Osservatorio Astronomico di Bologna, via Ranzani 1, I-40127 Bologna, Italy}

\date{Received \today; accepted ?}

\pagerange{\pageref{firstpage}--\pageref{lastpage}} \pubyear{2010}

\begin{document}
\label{firstpage}
\maketitle

\begin{abstract}
We study the imprints on the formation and evolution of cosmic structures of a particular class of dynamical dark energy models, characterized by an oscillating equation of state. This investigation complements earlier work on the topic, that focused exclusively on the expansion history of the Universe for such models. Oscillating dark energy cosmologies were introduced in an attempt to solve the coincidence problem, since in the course of cosmic history matter and dark energy would have had periodically comparable energy densities. In this class of models the redshift evolution of the equation of state parameter $w(z)$ for dark energy is characterized by two parameters, describing the amplitude and the frequency of the oscillations (the phase is usually set by the boundary condition that $w(z)$ should be close to $-1$ at recent times). We consider six different oscillating dark energy models, each characterized by a different set of parameter values. For one of these models $w(z)$ is lower than $-1$ at present and larger than $-1$ in the past, in agreement with some marginal evidence from recent type Ia supernova studies. Under the common assumption that dark energy is not clustering on the scales of interest, we study different aspects of cosmic structure formation. In particular, we self-consistently solve the spherical collapse problem based on the Newtonian hydrodynamical approach, and compute the resulting spherical overdensity as a function of cosmic time. We then estimate the behaviour of several cosmological observables, such as the linear growth factor, the Integrated Sachs-Wolfe (ISW) effect, the number counts of massive structures, and the matter and cosmic shear power spectra. We show that, independently of the amplitude and the frequency of the dark energy oscillations, none of the aforementioned observables show an oscillating behaviour as a function of redshift. This is a consequence of the said observables' being integrals over some functions of the expansion rate over cosmic history, thus smoothing any oscillatory features in $w(z)$ below detectability. We also notice that deviations with respect to the expectations for a fiducial $\Lambda$CDM cosmology are generically small, and in the majority of the cases distinguishing an oscillating dark energy model would be difficult. Exceptions to this conclusion are provided by the cosmic shear power spectrum, which for some of the models shows a difference at the level of $\sim 10\%$ over a wide range of angular scales, and the abundance of galaxy clusters, which is modified at the $\sim 10-20\%$ level at $z \gtrsim 0.6$ for future wide weak lensing surveys.
\end{abstract}


\section{Introduction}\label{sect:intro}

In recent years an increasingly large body of observations confirmed the general framework of a standard cosmological model based on General Relativity. Accordingly, right after the Big Bang the Universe experienced an accelerated expansion phase, dubbed \emph{inflation} \citep{Guth1981,Linde1982,Zelnikov1991}, during which quantum fluctuations were amplified to produce tiny perturbations in the matter distribution, whose imprints can nowadays be observed in the Cosmic Microwave Background (CMB) temperature map. Later on, and due to gravitational instability, these seed fluctuations grew up, giving rise to the web of cosmic structures that we observe today. After inflation, the Universe experienced a period of reheating \citep*{Shtanov1995,Kaiser1996} with the formation of light elements. Current observations of the CMB and of the luminosity distances of type Ia Supernovae (SNe Ia) show that the geometry of the Universe is spatially flat, in accordance with the predictions of the inflationary paradigm, and furthermore showing that the Universe is currently undergoing another accelerated expansion phase.

After the first detection of an accelerating expansion rate at low redshift (possibly $z\lesssim 0.5$, \citealt{Shapiro2006}) by \cite{Riess1998} and \cite{Perlmutter1999} many other different and independent studies led to the same conclusions making this inference very solid. In particular, evidence for an accelerated expansion comes from the CMB \citep{Jaffe2001,Komatsu2011} and the Integrated Sachs-Wolfe (ISW) effect \citep{Ho2008}, the Large Scale Structure (LSS) and the Baryon Acoustic Oscillation (BAO) \citep{Eisenstein2005,Percival2010}, globular clusters \citep{Krauss2003}, galaxy clusters \citep{Haiman2001,Allen2004,Allen2008,Wang2004} and weak lensing \citep{Hoekstra2006,Jarvis2006}.

As a homogeneous and isotropic General Relativistic model universe filled with matter is unable to reproduce the observed accelerated expansion, three different explanations have been proposed to account for it. One possibility consists in putting aside the hypothesis of homogeneity on large scales: these models are described by the Lema\^{\i}tre-Tolman-Bondi (LTB) metric or are based on the idea of backreaction \citep{Kolb2006}. A second possibility is to suppose that on very large scales General Relativity breaks down and gravity is modified. In this case we will be in the need for a new theory of gravity and General Relativity would be only the small-scale limit of a more profound theory. Examples of this idea are the $f(R)$ models \citep{Amendola2007,Starobinsky2007}, brane models \citep{Deffayet2001,Dvali2000}, and the $f(T)$ models \citep{Bengochea2009,Linder2010,Dent2011,Zheng2011}. Finally, one could assume that General Relativity is correct but the low-$z$ Universe is dominated by some kind of exotic fluid with negative pressure, the \emph{dark energy}. Specifically, if dark energy constitutes $\sim 70\%$ of the matter-energy content of the Universe, from the second Friedmann equation it turns out that its equation of state parameter $w$ would need to be $w<-1/2$ in order to provide accelerated expansion.

In the concordance cosmological model the role of dark energy is played by the cosmological constant $\Lambda$, having a redshift-independent equation of state parameter $w=-1$ and commonly interpreted as the energy density of the vacuum. Even though this $\Lambda$-Cold Dark Matter ($\Lambda$CDM henceforth) model is now the standard reference framework in cosmology, it suffers from some fundamental theoretical problems, that can be summarized by the following questions.

\begin{itemize}
\item Why is the energy density implied by the cosmological constant much smaller than the theoretically expected vacuum energy density?
\item Why is the dark energy density comparable to the dark matter density only today?
\end{itemize}
The last one is also known as the coincidence problem. In order to solve or at least alleviate these issues, it is possible to identify dark energy with the energy density of a minimally coupled scalar field (named \emph{quintessence}), that evolves through cosmic time as dictated by its own potential. This gives rise to a redshift-dependent equation of state parameter $w(z)$, hence making at least the coincidence problem less severe. These dynamical dark energy models can be roughly grouped into two classes, tracking models \citep{Steinhardt1999} and scaling models \citep{Halliwell1987, Wands1993, Wetterich1995}.

Models with an oscillating equation of state were introduced to solve the coincidence problem, because the present accelerated expansion phase would just be one of the many such phases occurring over cosmic history, especially at early times. Moreover, oscillations would more naturally accommodate the crossing of the phantom barrier, $w=-1$ as it is marginally suggested by recent observations \citep{Alam2004,Allen2004,Dicus2004,Riess2004,Feng2005,Huterer2005,Choudhury2005}. In the framework of particle physics, it is possible to have an oscillating quintessence potential if one considers a pseudo-Nambu-Goldstone boson field when it has rolled through the minimum. As models for dark energy, oscillating scalar fields were proposed by \cite{Dutta2008}, \cite{Johnson2008}, and \cite{Gu2008}. An oscillating behaviour can also be obtained in models with growing neutrino mass, where the dark energy component is coupled with massive neutrinos. These models in fact predict oscillations in the dark energy equation of state for relatively low redshifts ($z\lesssim 10$), see e.g. \cite*{Amendola2008}, \cite{Mota2008a}, \cite{Wintergerst2010}, \cite{Baldi2011}. \cite{Lazkoz2005} and \cite{Kurek2008} found a better agreement with SNe Ia data if an oscillating equation of state is used instead of the cosmological constant or an equation of state linearly dependent on the scale factor. Further indications in the same direction come from the study performed by \cite{Riess2007}. Previous works on this topic focused mainly on the expansion history of the Universe and marginally on the linear perturbation theory in the framework of oscillating quintessence \citep{Feng2005,Xia2005,Barenboim2006a,Barenboim2006b,Kurek2008,Kurek2010,Lan2010}.

The novelty of this work lies in the fact that we explore signatures of an oscillating equation of state $w(z)$ in the (non-linear) growth of cosmic structures, thus extending and complementing the majority of foregoing studies. The main idea we explore here is to find out cosmological observables based on structure formation that can hint toward oscillating quintessence even though the expansion of the homogeneous and isotropic background does not. The rest of this paper is hence organized as follows. In Section~\ref{sect:spc} we describe the formalism of the spherical collapse used to derive important parameters for the formation and evolution of structures. In Section~\ref{sect:models} we describe and motivate several parametrizations used in order to describe the oscillating dark energy and in Section~\ref{sect:results} we present results for the different observables we considered. Section~\ref{sect:conclusions} is devoted to our conclusions. In Appendix~\ref{sect:code} we present implementation details of the code used to evaluate the linear growth factor and the evolution of the spherical overdensity.

\section{Spherical collapse model}\label{sect:spc}

Despite its simplifying nature, the model describing the collapse of a uniform non-rotating spherical overdensity in a cosmological setting provides numerous insights on the actual process of structure formation. For instance, the linear density contrast extrapolated at the spherical collapse time provides a fair approximation for the threshold at which actual perturbations can collapse to form bound structures. Thus, in this Section we sketch the derivation of the relevant equations for the spherical collapse model under the assumption that only dark matter can form clumps, while dark energy is just present as background fluid. For further details we refer to the current literature on the topic \citep[see, e.g.][]{Bernardeau1994, Ohta2003, Ohta2004, Mota2004, Nunes2006, Abramo2007, Pace2010}.

Rather than studying the time evolution of the radius of the collapsing sphere, we study directly the time evolution of the overdensity. This procedure proves to be numerically more stable and less prone to errors than the classical approach based on the radius evolution \citep{Pace2010}. We consider a perfect fluid described by the energy-momentum tensor $T^{\mu\nu}$, satisfying the local conservation laws expressed by $\nabla_{\nu} T^{\mu\nu}=0$. This set of four equations encapsulates both the continuity and the Euler equations, while from Einstein's field equations it is possible to derive a relativistic generalization of the Poisson equation. In a more explicit form these expressions read

\begin{equation}\label{eqn:cnpert}
\frac{\partial\rho}{\partial t}+\vec{\nabla}\cdot(\rho\vec{v})+\frac{P}{c^2}\nabla\cdot\vec{v} =  0~,
\end{equation}
 
\begin{equation}\label{eqn:enpert}
 \frac{\partial\vec{v}}{\partial t}+(\vec{v}\cdot\vec{\nabla})\vec{v}+\vec{\nabla}\Phi =  0~,
\end{equation}
and

\begin{equation}
 \nabla^2\Phi-4\pi G\left(\rho+\frac{3P}{c^2}\right) =  0~,
\end{equation}
where $\rho$, $P$, $\vec{v}$ and $\Phi$ are the density, the pressure, the velocity and the gravitational potential of the fluid.\\
For the average background matter density the following continuity equation holds,

\begin{equation}
 \dot{\bar{\rho}}+3H\left(\bar{\rho}+\frac{\bar{P}}{c^2}\right)=0~,
\end{equation}
where $\bar{\rho}=3H^2\Omega_{\mathrm{m}}/8\pi G$ is the background matter density, $H$ is the Hubble parameter, and $\Omega_\mathrm{m}$ is the matter density parameter. Since for ordinary matter and dark matter the pressure contribution is negligible, from now on we will set $P=0$.

Assuming spherical symmetry and perturbing the physical quantities appearing in the previous set of equations (density, velocity and gravitational potential) around their background values, we obtain the following exact non-linear differential equation, describing the evolution of the matter density perturbation $\delta$ as a function of the cosmic time,

\begin{equation}\label{eqn:nleq}
 \ddot{\delta}+2H\dot{\delta}-\frac{4}{3}\frac{\dot{\delta}^2}{1+\delta}-4\pi G\bar{\rho}~\delta(1+\delta)=0~.
\end{equation}
We stress that Eq. (\ref{eqn:nleq}) is valid also for large density contrasts, deep in the non-linear regime, as long as spherical symmetry is satisfied. By restricting to $\delta \ll1$ instead, at first order Eq. (\ref{eqn:nleq}) reads

\begin{equation}\label{eqn:leq}
 \ddot{\delta}+2H\dot{\delta}-4\pi G\bar{\rho}~\delta=0~,
\end{equation}
and it coincides with the differential equation commonly used to determine the linear growth factor.

As explained in detail in \cite{Pace2010}, in order to determine the linear density perturbation threshold for spherical collapse, $\delta_\mathrm{c}$, one should solve Eq. (\ref{eqn:leq}) with suitable initial conditions, namely the initial overdensity and velocity of the perturbation. In order to find the initial overdensity we take into account that at the time of the collapse of the object all the matter is concentrated in one point, therefore formally $\delta\rightarrow+ \infty$. Hence by fixing the time, or scale factor, of collapse $a_{\mathrm{c}}$, with a root-search method it is possible to determine the initial overdensity $\delta_{\mathrm{i}}$ such that the solution of the non-linear Eq. (\ref{eqn:nleq}) diverges at $a_{\mathrm{c}}$. Once the initial overdensity is found, we compute the initial velocity as detailed in Appendix \ref{sect:code}, and use both of them as initial conditions for the linear Eq.~(\ref{eqn:leq}). When integrated up to $a_\mathrm{c}$ the latter returns us the linear density contrast corresponding to the time of spherical collapse, $\delta_\mathrm{c}$.

In order to determine the virial overdensity $\Delta_{\mathrm{v}}$, representing the non-linear evolution of the density perturbation up to the time of virialization, we need to evaluate the turn-around scale factor $a_{\mathrm{ta}}$, defined as the time when the radius of the sphere reaches its maximum, detaches from the overall expansion of the Universe, and collapses afterwards. Using then the virial theorem and energy conservation considerations, the virial overdensity $\Delta_{\mathrm{v}}$ can be derived according to the discussion of \cite{Maor2005}.

\section{Outline of the cosmological models}\label{sect:models}

As outlined in Section \ref{sect:intro}, at the moment there is no explanation for dark energy in terms of fundamental physics, therefore all the models that we explored in the present work are purely phenomenological and the values of their parameters are generically adjusted such that certain classes of cosmological observables (most commonly the luminosity distance of SNe Ia and the CMB temperature power spectrum) are well reproduced. These models are described in the present Section.
As a fiducial reference cosmology we assume  the standard flat $\Lambda$CDM model. The cosmological parameters are set to $\Omega_{\mathrm{m},0}=0.274$, $\Omega_{\mathrm{q},0}=0.726$ and $h=0.7$ in accordance with WMAP-7 data \citep{Komatsu2011,Larson2011} and the Supernova Legacy Survey 3 data \citep[see][]{Sullivan2011}. These same parameter values are also kept intact for all the dynamical dark energy cosmologies investigated in this paper.

The amplitude of the primordial matter power spectrum in the fiducial $\Lambda$CDM cosmology is selected in order to attain a given value of the quadratic deviation on a comoving scale of $8$ Mpc/$h$, $\sigma_8=0.8$. In all the other dynamical dark energy models that we considered the normalization is scaled according to 

\begin{equation}
\sigma_{8,\mathrm{DE}}=\frac{\delta_{\mathrm{c,DE}}(z=0)}{\delta_{\mathrm{c,\Lambda CDM}}(z=0)}\sigma_{8,\Lambda\mathrm{CDM}}\;,
\end{equation}
where $\delta_{\mathrm{c}}$ is the linear overdensity parameter extrapolated at spherical collapse (see Section \ref{sect:spc} above). In this way, the exponential tail of the dark matter halo mass function at redshift zero is conserved, and hence the abundance of massive structures at present times, which is arguably well defined from the observational point of view, is the same for all models. We show the values for the normalization of the different models in Table~\ref{tab:eos_param}.

In the present work we analysed six different dark energy cosmologies with an oscillating equation of state parameter $w(z)$. In the first five of them $w(z)$ has the same functional form but different values of the free parameters. The sixth model has instead a different functional form for $w(z)$, although still presenting oscillations. The functional form for the first five cosmologies is \citep{Linder2006,Lakzok2010,Feng2006}

\begin{equation}
w(a)=w_0-A\sin{(B\ln{a}+\theta)}~,
\end{equation}
where $a=1/(1+z)$ is the scale factor, $A$ determines the amplitude of the oscillations, $B$ gives their frequency while $\theta$ is a phase shift. As can be easily seen, the value of the equation of state parameter today is $w(a=1) = w_0-A\sin (\theta)$, which equals $w_0$ if the phase is $\theta=0$. Model six is meant to generalize the CPL parametrization \citep{Chevallier2001,Linder2003} in order to avoid the future unphysical divergence of the dark energy equation of state typical of this model. In this case, the function $w(z)$ can be written as

\begin{equation}
 w(a)=w_0-A(aB\sin{(1/a)}+\theta)~.
\end{equation}
Going into the distant future, we have $a\rightarrow\infty$ and taking the limit of the previous equation we observe that it asymptotes to

\begin{equation}
 w(a)=w_0-A(B+\theta)~,
\end{equation}
thus remaining finite. For more details on this model we refer the reader to \cite{Ma2011}.

In Table~\ref{tab:eos_param} we summarize the values for the free parameters characterizing each cosmology. We also quote the paper where the expansion history of the Universe resulting from that specific dark energy model has been studied in detail. The quoted values are the best fit to certain classes of cosmological observables considered by those authors. In the literature other parametrizations of $w(z)$ showing an oscillatory behaviour can be found \citep[see for example][]{Kurek2008,Kurek2010}. However they represent only local fits to the expansion history, giving unphysical divergences in the distant past. For this reason we did not include them in our analysis.

\begin{table}
\caption{Values of the free parameters for the dark energy equation of state and for the matter power spectrum normalization $\sigma_{8}$. The numbers in the last column correspond to the following references: $(1):$ \citet{Linder2006}, $(2):$ \citet{Lakzok2010}, $(3):$ \citet{Feng2006}, $(4):$ \citet{Ma2011}.}
\label{tab:eos_param}
\begin{center}
\begin{tabular}{c|c|c|c|c|c|c}
\hline
\hline
Model & $w_0$ & $A$ & $B$ & $\theta$ & $\sigma_{8}$ & Reference\\
\hline
1 & -0.9 & 0.07 & 5.72 & 0.0 & 0.7989 & (1) \\
2 & -0.9 & 0.07 & 2.86 & 0.0 & 0.7986 & (1) \\
3 & -0.9 & 0.15 & 1.0 &  0.0 & 0.7983 & (1) \\
4 & 0.0 & 1.0 & 0.06 & $\pi/2$ & 0.7999 & (2) \\
5 & -1.0 & 1.5 & 0.032 & $5\pi/18$ & 0.8012 & (3) \\
6 & -1.061 & 0.041 & 1.0 & $-\sin(1)$ & 0.8001 & (4)\\
\hline
\hline
\end{tabular}
\end{center}
\end{table}

\begin{figure}
\includegraphics[angle=-90,width=\hsize]{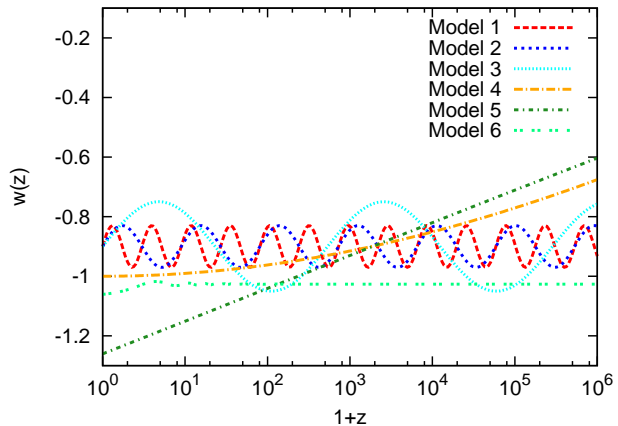}
\caption{The redshift evolution of the equation of state parameter for the oscillating dark energy cosmologies analysed in this work. The red dashed, blue short-dashed and cyan dotted curves show the models 1, 2 and 3 respectively. The orange dot-dashed curve shows model 4, while the dark-green dot-short-dashed and the light-green dot-dotted lines represent models 5 and 6, respectively (see Table \ref{tab:eos_param} for more details).}
\label{fig:wz}
\end{figure}

In Figure~\ref{fig:wz} we show the redshift evolution of the equation of state parameter $w(z)$ for the different models studied in this work. We refer to the caption for the description of the line styles adopted. As it appears evident from the Figure, models 1 and 2 have the same amplitude but the frequency of oscillations changes by a factor of two between each other. Model 3 is qualitatively the same, but the amplitude is twice as big and the oscillations one third less frequent with respect to model 1. These models allow a comparative study on the influence of the amplitude and the frequency of the oscillations. Models 4 and 5, despite being described by the same functional form, have a very long period and oscillations are not even visible throughout cosmic history. Models 3 and 5 are also characterized by a crossing of the phantom barrier, $w = -1$, a feature that is marginally find to best fit the luminosity distance of SNe Ia. Finally, model 6 shows tiny oscillations only at recent times, while for $z\gtrsim 10$ the function $w(z)$ approaches a constant. By comparing model 6 with the others we can draw conclusions about the importance of the oscillations at early times.

\subsection{Homogeneous background analysis}

\begin{figure}
\includegraphics[angle=-90,width=0.45\textwidth]{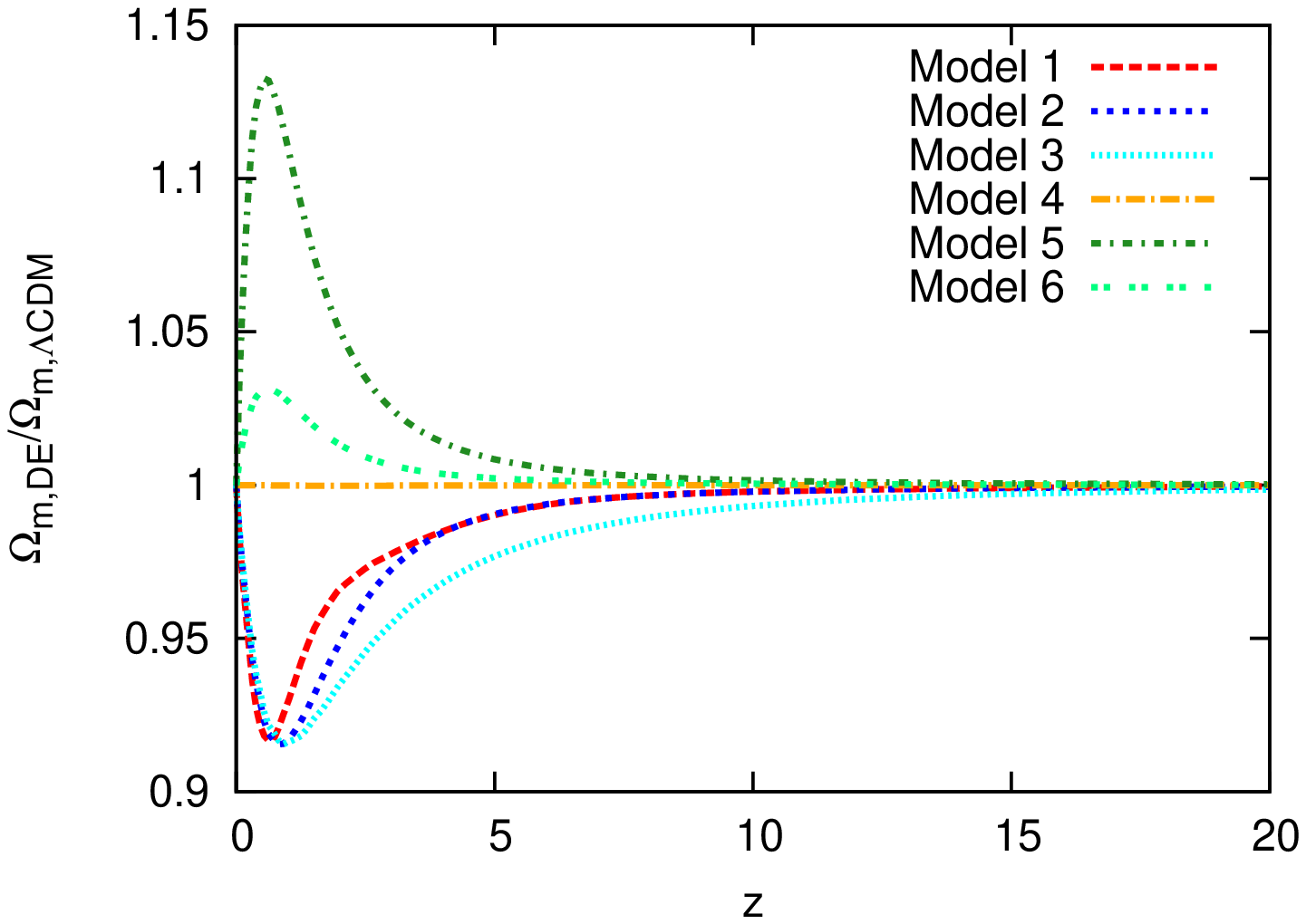}
\includegraphics[angle=-90,width=0.45\textwidth]{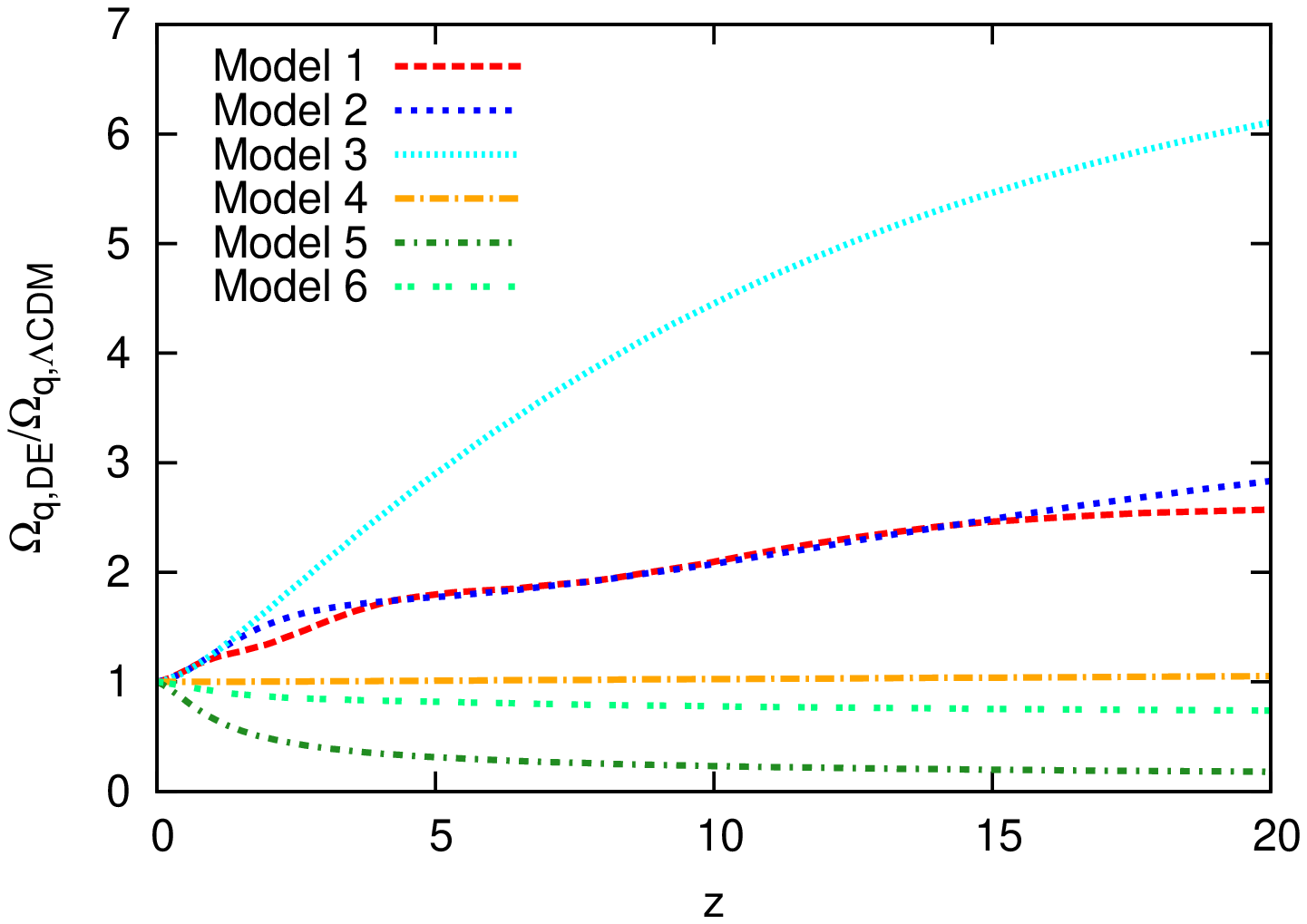}
\caption{The redshift evolution of the density parameters. In the upper (lower) panel we show the ratio of the matter (dark energy) density parameter in the six oscillating dark energy cosmologies studied in this work to the corresponding function in the fiducial $\Lambda$CDM model. Line types and colours are as in Figure~\ref{fig:wz}.}
\label{fig:omega}
\end{figure}

We now explore in detail the redshift evolution of functions related to the homogeneous background for the various oscillating quintessence cosmologies presented above. In the upper panel of Figure~\ref{fig:omega} we show the ratio between the matter density parameters in the six dynamical dark energy models considered here and the same function in the concordance $\Lambda$CDM cosmology, as a function of redshift. In the lower panel of the same Figure we display the corresponding redshift evolution for the dark energy density parameters. As one could naively expect, the amount of matter at early times is the same in all models, consequence of the fact that at high redshift the dark energy contribution becomes negligible, and hence the Universe always behaves as an Einstein-de Sitter (EdS) cosmology. Differences become significant at $z \lesssim 5$ and are at most at the level of $\sim 10-15\%$. It is worth noticing that in no circumstances we see full oscillations in the density parameters, implying that one integral over the cosmic history is enough to smooth out most features of $w(z)$. 

It is worth noting that model 6 and, especially, model 4 display very little difference with respect to the concordance case, an instance that will show up time and again throughout the discussion of our results. The other models instead, with the exception of model 5 that sees a reduction in the abundance of dark energy, show a substantially higher amount of dark energy at early times than the cosmological constant case. At $z\sim 20$ the difference in the dark energy density parameter is of one order of magnitude or more. It should be recalled however that at such high redshifts the contribution of dark energy to the expansion history of the Universe is negligible anyway.

\begin{figure}
\includegraphics[angle=-90,width=0.45\textwidth]{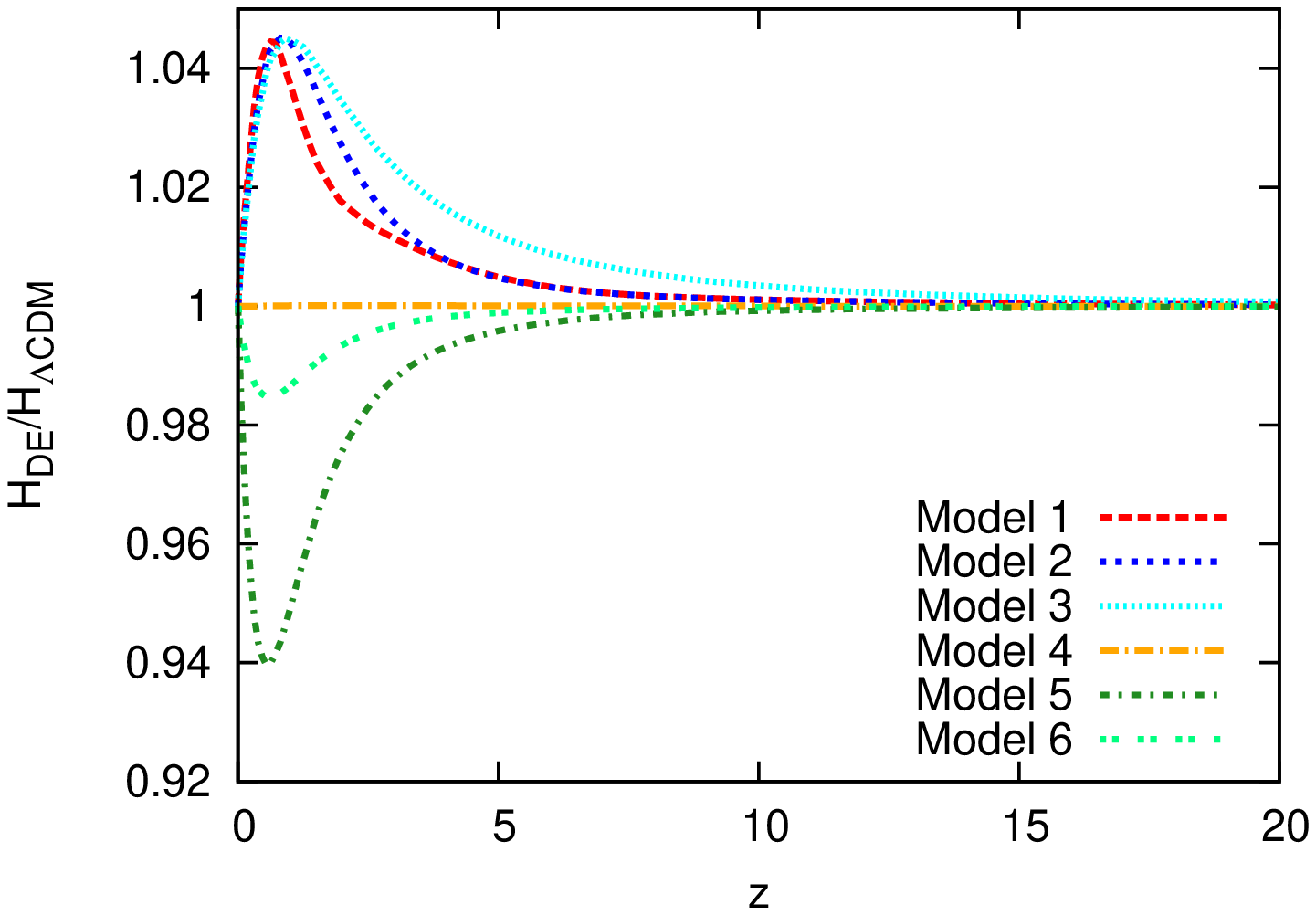}
\includegraphics[angle=-90,width=0.45\textwidth]{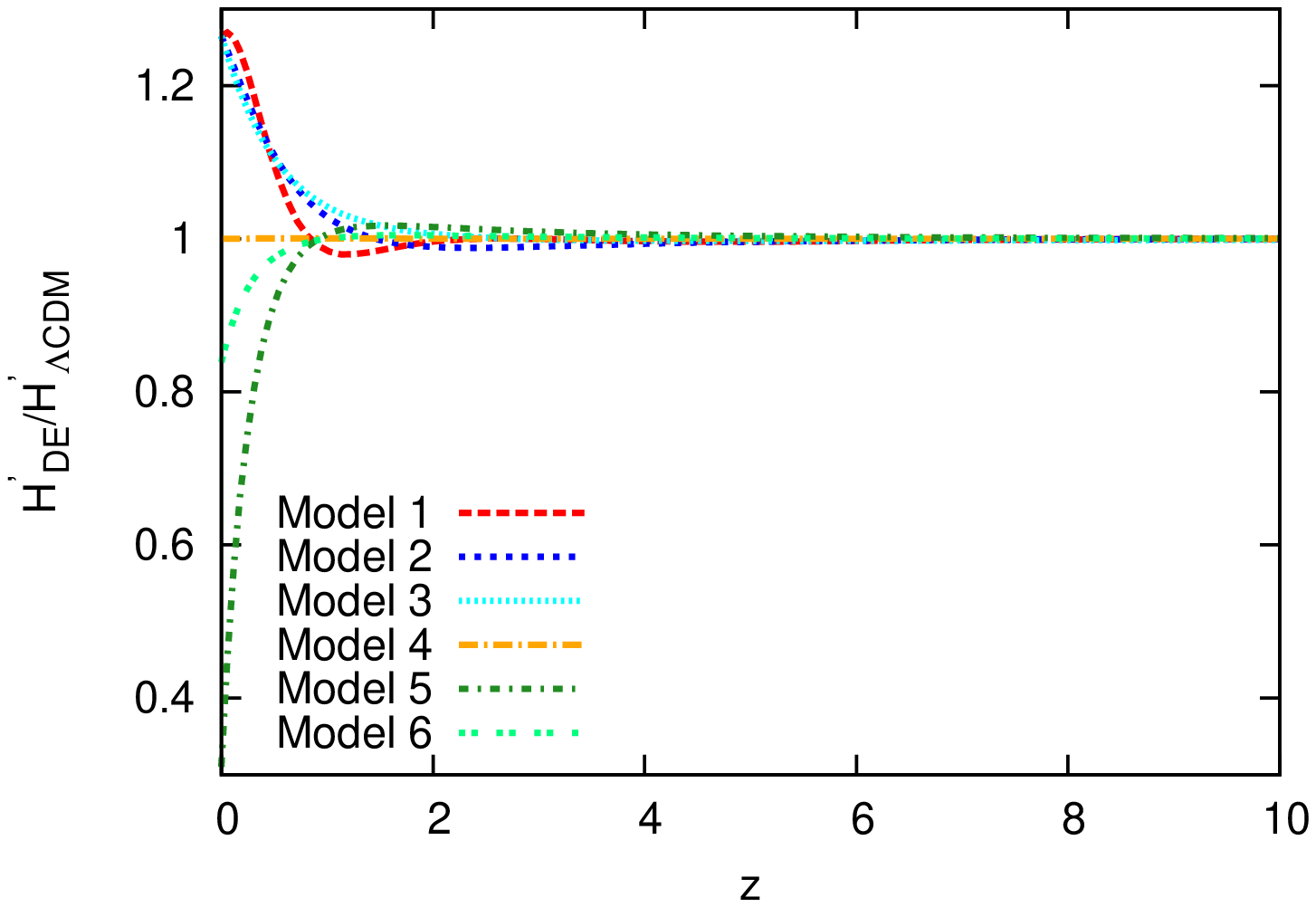}
\caption{\emph{Upper (lower) panel.} Ratio of the Hubble parameter (derivative of the Hubble parameter with respect to the scale factor) in the six oscillating dark energy models considered in the present paper to the same function in the fiducial $\Lambda$CDM cosmology. Line styles and colours are as in Figure~\ref{fig:wz}.}
\label{fig:hubble}
\end{figure}

A different perspective on the same results is given by examining the Hubble parameters (that are, the expansion rates) for the various cosmologies, shown in the upper panel of Figure~\ref{fig:hubble}. For flat universes, the Hubble parameter can be written as

\begin{equation}
H(a)=H_0E(a)=H_0\sqrt{\frac{\Omega_{\mathrm{m},0}}{a^3}+\Omega_{\mathrm{q},0}~g(a)}~,
\end{equation}
where $g(a)$ is defined as 
\begin{equation}\label{eqn:g}
 g(a)=\exp{\left(-3\int_1^a \frac{1+w(a^\prime)}{a^\prime}da^\prime\right)}\;,
\end{equation}
and $\Omega_\mathrm{q}$ is the dark energy density. The behaviour of the expansion rates is very similar to that of the matter density parameters, which is expected because dark energy comes to dominate the evolution of the Hubble function only at very low redshift, where differences between different models tend to vanish.

Since oscillations are not present in the redshift evolution of the matter and dark energy density parameters, the same holds true for the expansion rate in our quintessence cosmologies. However, considering the derivative of the expansion rate with respect to the scale factor (dubbed \emph{deceleration} parameter, see \cite{Dunajski2008} and references therein), shown in the lower panel of Figure \ref{fig:hubble}, we observe some partial indication of oscillations, in that the oscillating dark energy deceleration parameter crosses the $\Lambda$CDM one at least once for models 1, 2, and 5. However, since the overall pattern looks the same for all models, it is likely not directly connected with dark energy oscillations. Moreover, oscillating quintessence models introduce absolute differences in the deceleration parameter of at least $\sim 20-30\%$ with respect to the fiducial case, which is a quite significant effect.

\begin{figure}
\includegraphics[angle=-90,width=0.45\textwidth]{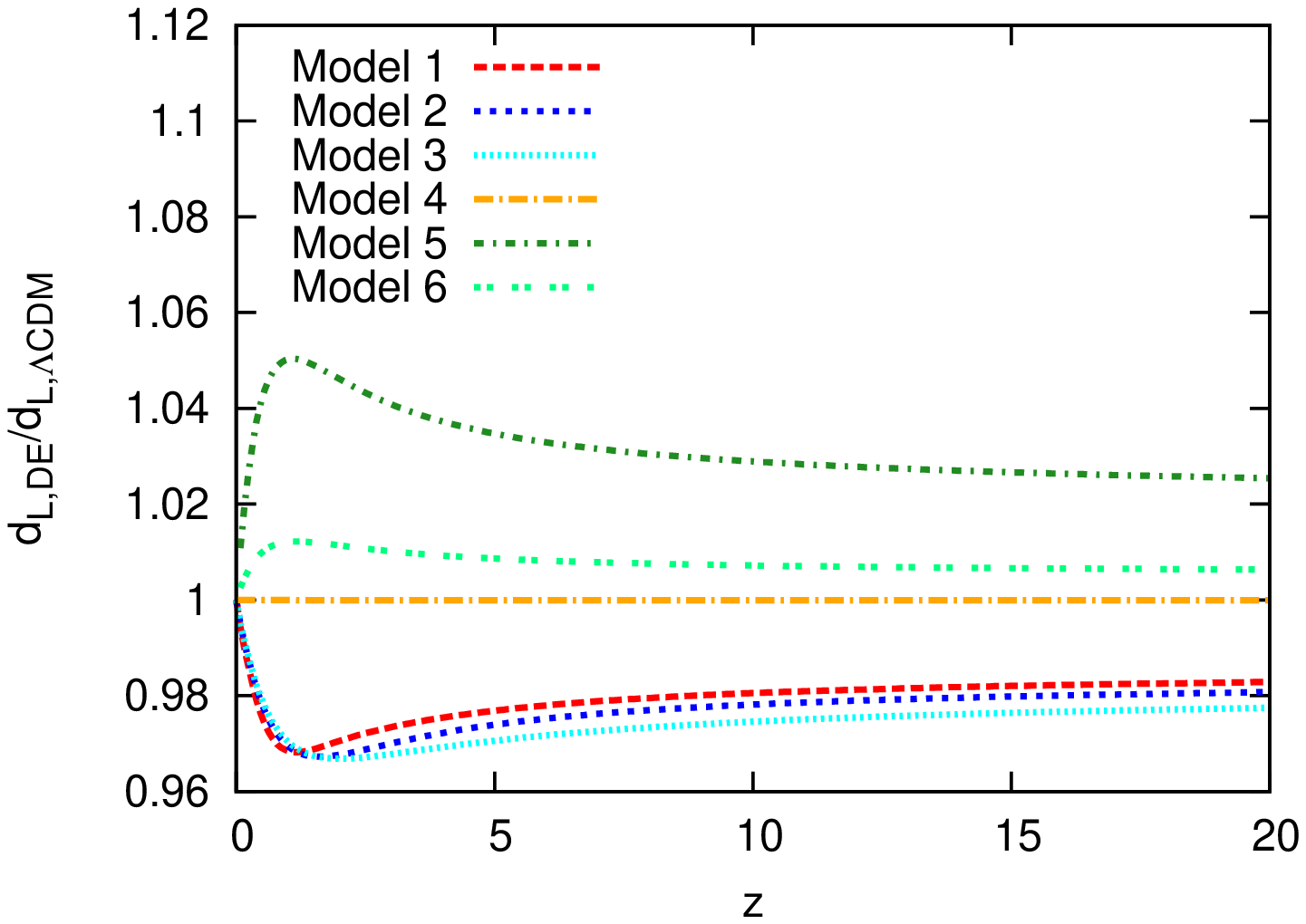}
\includegraphics[angle=-90,width=0.45\textwidth]{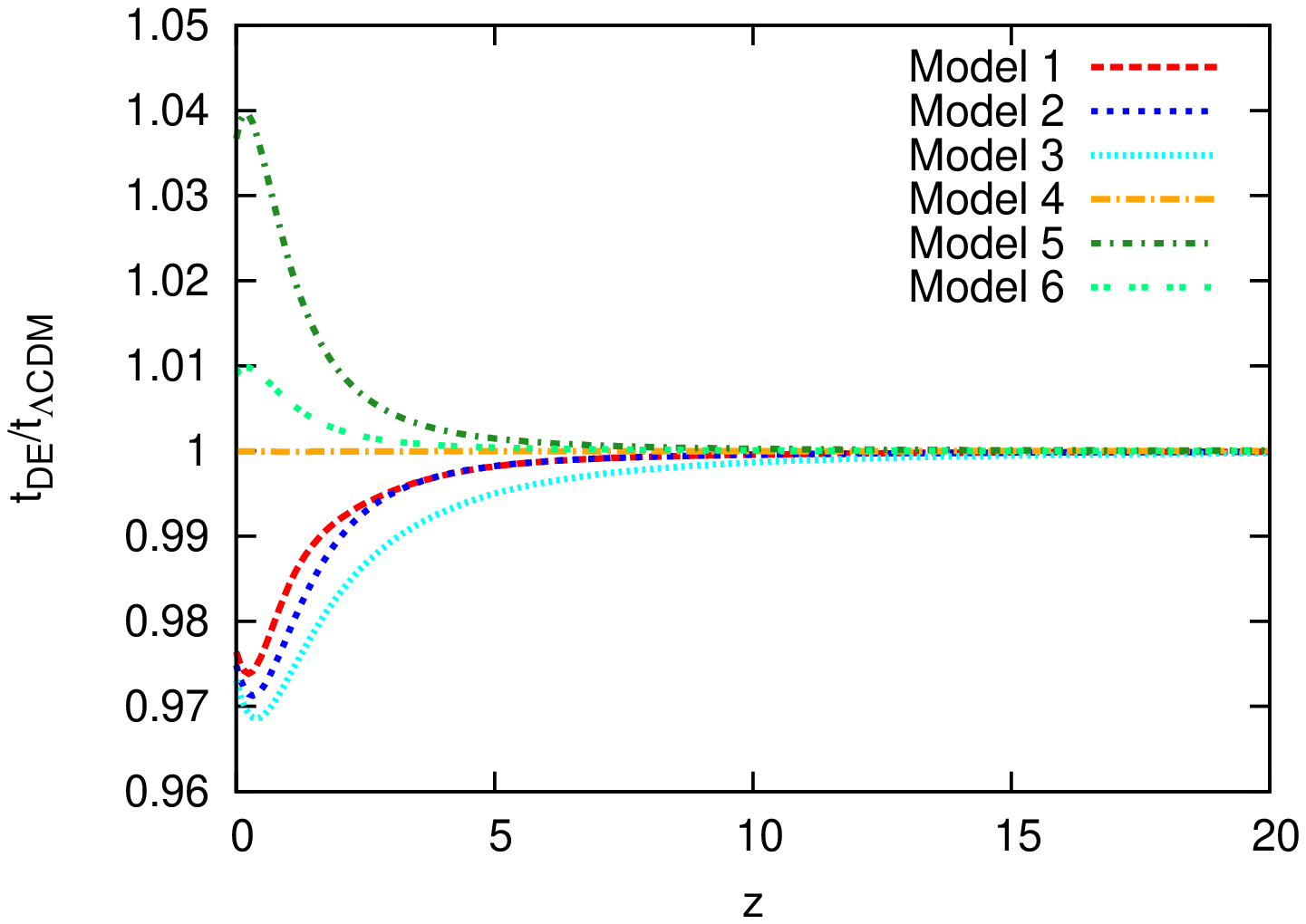}
\caption{\emph{Upper (lower) panel.} Ratio of the luminosity distances (ages of the Universe) in the oscillating dark energy models considered here to the same quantity for the fiducial $\Lambda$CDM cosmology, as a function of redshift. Line styles and colours are as in Figure~\ref{fig:wz}.}
\label{fig:dist_time}
\end{figure}

In the upper (lower) panel of Figure~\ref{fig:dist_time} we present the ratio of the luminosity distance (age of the Universe) in the various dynamical dark energy cosmologies to the same quantity in the $\Lambda$CDM model, as a function of redshift. In both cases we see that differences are at most of the order of $\sim 4\%$ and predominantly located at relatively low redshifts ($z \lesssim 2$), although the luminosity distance shows a $\sim 2\%$ deviation even at arbitrarily high redshifts. This fact is expected, since both the luminosity distance and the age of the Universe are suitable integrals over some function of the Hubble parameter, which also shows most differences at low redshift. Comparing the upper panel of Figure~\ref{fig:dist_time} with the luminosity distances inferred by SN Ia Union2 data \citep{Amanullah2010} we see that differences induced by the oscillating dark energy models at low redshifts are at the same level of the systematic errors in the measurements as well as of the intrinsic scatter around the best fit. This, together with the very slight deviations in various cosmological functions shown in previous Figures lead us to conclude that the oscillating quintessence cosmologies studied in this work are not distinguishable from the concordance model by current geometrical probes. This is perfectly consistent with previous works, since the parameter values that we adopted are indeed chosen so as to reproduce some particular geometrical tests.

\begin{figure}
\includegraphics[angle=-90,width=0.45\textwidth]{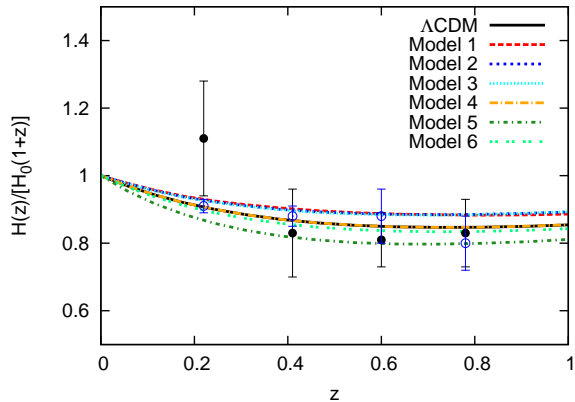}
\caption{Redshift behaviour of the quantity $H(z)/[H_{0}(1+z)]$ for the $\Lambda$CDM model and the six oscillating dark energy models considered in this work. The black solid line represents the concordance $\Lambda$CDM cosmology, the other models are shown using the same line styles and colours as in Figure~\ref{fig:wz}. Black and blue data points with error bars are from the WiggleZ Dark Energy Survey.}
\label{fig:ez_points}
\end{figure}

It is also interesting to compare our models with recent measurements of the Hubble function performed with the WiggleZ Dark Energy survey \citep{Drinkwater2010}. These measurements, together with the determination of the growth rate (see Sect.~\ref{sect:gf}), represent the most accurate and the highest redshift ones available at the moment. This comparison is done in Figure~\ref{fig:ez_points}, where we plot the quantity $H(z)/[H_{0}(1+z)]$ for the $\Lambda$CDM model and the six oscillating models considered in this work. The black data points (filled circles with error bars) are obtained using the Alcock-Paczynski \citep{Alcock1979} test in combination with SNIa distance measurements. The blue points (open circles) are results of the same test, but obtained with the distance reconstruction method of \cite{Shafieloo2006}. Notice how this method is able to dramatically reduce error bars. Data points are taken from \cite{Blake2011d} to which we refer the reader for more details on the analysis leading to their determination of the four data points. The first thing to notice is that our $\Lambda$CDM curve would be a slightly worse fit to the data points with respect to Figure~{5} from \cite{Blake2011d} since we use slightly different cosmological parameters. This being said, we observe that, despite the very good quality of the data, the size of the 1-$\sigma$ error bar is too large to rule out the oscillating models considered in this work. All the models are consistent within 3-$\sigma$ with results inferred from observations. We can therefore safely use these models for the following analysis.

\subsection{Redshift drift}

An important cosmological test related with the expansion history that has not been considered in the past, but might reveal itself valuable in the near future is the so called \emph{redshift drift}, that represents the variation of the cosmological redshift of a source due to the expansion of the Universe \citep{Balbi2007,Liske2008,Uzan2008,Jain2010,Araujo2010}. Let us indicate with $t_{\mathrm{s}}$ the time of emission of an electromagnetic signal from a source, and with $t_0$ the time of observation of the same signal. The cosmological redshift of the source is then defined as

\begin{equation}
1+z_{\mathrm{s}}=\frac{a(t_{0})}{a(t_{\mathrm{s}})}\;.
\end{equation}
After a time interval $\Delta t_{0}$ has passed for the observer, corresponding to an interval $\Delta t_{\mathrm{s}}$ for the source, the change in the source redshift can be estimated by expanding at first order the previous equation,

\begin{equation}
\Delta z_{\mathrm{s}}\simeq\Delta t_{0}\left[\frac{\dot{a}(t_{0})-\dot{a}(t_{\mathrm{s}})}{a(t_{\mathrm{s}})}\right]\;.
\end{equation}
By substituting $H(a)=\dot{a}/a$ we obtain the expression for the redshift drift

\begin{equation}
\Delta z_{\mathrm{s}}\simeq H_{0}\Delta t_{0}\left[1+z_{\mathrm{s}}-E(z_{\mathrm{s}})\right]\;,
\end{equation}
where $E(z_{\mathrm{s}})=H(z_{\mathrm{s}})/H_{0}$.

Using the variation in the cosmological redshift, it is also possible to determine the variation in the recession velocity of the source,

\begin{equation}
\Delta\upsilon_\mathrm{s}=\frac{c\Delta z_{\mathrm{s}}}{1+z_{\mathrm{s}}}\;.
\end{equation}
We can therefore write

\begin{equation}
\dot{\upsilon}_\mathrm{s}=\frac{cH_{0}}{1+z_\mathrm{s}}[1+z_\mathrm{s}-E(z_\mathrm{s})]\;.
\end{equation}
Since all the cosmological properties of the model at hand are encoded into the Hubble function, we see that we can use the time variation of the redshift in order to reconstruct the expansion history of the Universe.

\begin{figure}
\includegraphics[angle=-90,width=0.45\textwidth]{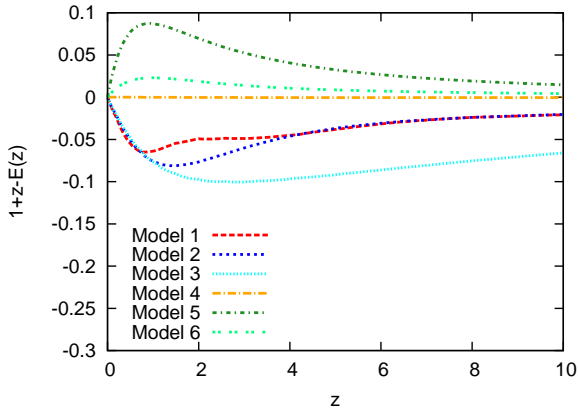}
\caption{The difference between the redshift drifts in the six oscillating quintessence cosmologies considered in this work and the some quantity in the fiducial $\Lambda$CDM model, as a function of redshift. Line styles and colours are the same as in Figure~\ref{fig:wz}.}
\label{fig:redshift_drift}
\end{figure}

In Figure~\ref{fig:redshift_drift} we show the difference between the redshift drifts in the oscillating dark energy models and the same function in the concordance $\Lambda$CDM cosmology, per unit of observed time and normalized by the Hubble constant. We decided to plot the differences instead of ratios in this case in order to avoid divergences, due to the fact that $\Delta z_{\mathrm{s}}$ goes to zero when $1+z_\mathrm{s} = E(z_\mathrm{s})$. Similarly to previous Figures, also in this case models 4 and 6 behave very similarly to the $\Lambda$CDM cosmology. The redshift drift is systematically higher for the model 5 while it is systematically lower for the first three oscillating cosmologies. At high redshifts the differences between the models decrease since all the Hubble parameters converge to the EdS behaviour. Since all the cosmological information is encoded in the Hubble expansion function, no oscillations appear in this case as well, although a slight wiggle is visible for model 1 at $z\sim 2$. As for the perspective of realistically measuring the redshift drift, according to \cite{Liske2008}, peculiar motions are negligible ($\sim 10^{-3}$~cm s$^{-1}$) and with a temporal baseline of 20 years it will be possible to determine the existence of the cosmological constant at $3.1~\sigma$ observing distant quasars. Following \cite{Balbi2007} we also notice that variations in the recession velocity of the sources are bigger than the error bars forecasted by Montecarlo simulations, therefore with a sufficiently long baseline, it should be possible to discriminate between oscillating quintessence and cosmological constant, at least for the models showing more significant deviations.

\section{Results}\label{sect:results}

In this Section we present results concerning the structure formation in the oscillating quintessence cosmologies described in Section~\ref{sect:models}. We studied several aspects of structure formation and in particular we focused our attention on the growth factor, the linear and non-linear overdensities derived from the spherical collapse model, the mass function of cosmic structures, the power spectrum of cosmic shear and the ISW effect. We now proceed to describe each one of these observables in detail.

\subsection{Growth factor}\label{sect:gf}

\begin{figure}
\includegraphics[angle=-90,width=0.45\textwidth]{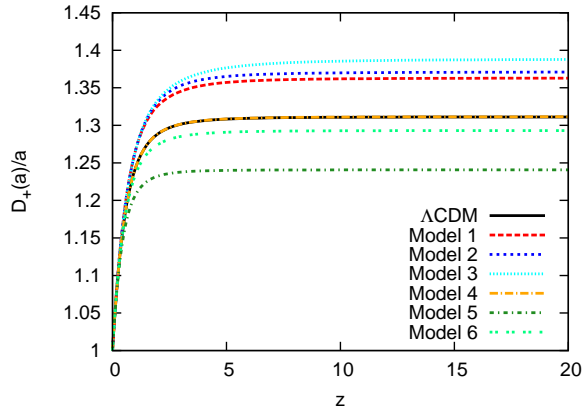}
\caption{The growth factor as a function of the redshift. Line styles and colours are as in Figure~\ref{fig:ez_points}.}
\label{fig:gf}
\end{figure}

In Figure~\ref{fig:gf} we show the growth factor normalized by the scale factor, $D_{+}(a)/a$, as a function of redshift for the six oscillating dark energy cosmologies described above, plus the fiducial $\Lambda$CDM model. Cosmological observables sensitive to the growth factor include cosmic shear, ISW effect and the Rees-Sciama effect, all discussed later on. In Figure \ref{fig:gf} the growth factor is normalized to unity at $z=0$. As can be seen, differences between the oscillating quintessence models and the $\Lambda$CDM cosmology (solid black curve) are at most of $\sim 10\%$ (for model 3), while, in agreement with previous Figures, model 4 does not show any appreciable difference from the concordance model. It is interesting to note that, while $w(z)$ for model 4 shows indeed very little variation up to the last scattering redshift due to the very large period of its oscillations (see Figure~\ref{fig:wz}), model 6 shows even smaller time evolution, yet its effects on cosmological functions are larger. This implies that high frequency oscillations in $w(z)$, albeit with a very small amplitude and limited time extent have more of an effect on the expansion history (and structure formation too, see later Subsections) than larger oscillations with a low frequency. This is because low-frequency oscillations cancel integral contributions more effectively.

Figure~\ref{fig:gf} also shows that for models 5 and 6 the growth factor is smaller than for the cosmological constant case. This can be understood by the following argument. From the lower panel of Figure~\ref{fig:omega} we see that for these two models the amount of dark energy is smaller than for the cosmological constant case at all redshifts. This means that the Hubble drag is less effective in the former models, and hence the growth of structures (at least at the linear stage) is easier. Since the amplitude of density fluctuations at $z = 0$ is almost the same amongst all the dark energy models considered here (it differs by a factor proportional to the critical overdensity for spherical collapse, that however is only slightly changed in the case of dynamical dark energy, see below), the growth factor must be smaller in order to match the amplitude of fluctuations at early times.

The growth factors depicted in Figure \ref{fig:gf} do not present any sign of oscillations, not even if we consider their ratios with respect to the concordance cosmology case. By rewriting Eq.~(\ref{eqn:leq}) using the scale factor instead of cosmic time as the independent variable, we obtain

\begin{equation}
\delta^{\prime\prime}+\left(\frac{3}{a}+\frac{E^\prime}{E}\right)\delta^\prime-\frac{3}{2}\frac{\Omega_{\mathrm{m},0}}{a^5E^2}\delta=0~.
\end{equation}
As shown above, the derivative of the Hubble function presents mild signs of oscillations, however since this is a second order differential equation, the solution involves a double integral over the scale factor, that efficiently smoothes out any fluctuation in the coefficients.

\begin{figure}
\includegraphics[angle=-90,width=0.45\textwidth]{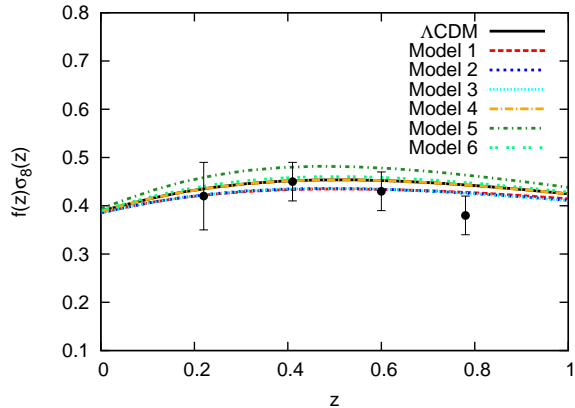}
\caption{Growth rate of structure as a function of redshift, expressed in a more model-independent way via the function $f(z)\sigma_{8}(z)$. Black data points are from the WiggleZ Dark Energy Survey. Line styles and colours are as in Figure~\ref{fig:ez_points}.}
\label{fig:fsigma8}
\end{figure}

Finally, we also estimated the logarithmic derivative of the growth factor with respect to the scale factor, $f(\Omega_\mathrm{m}(a))=d\ln{D_{+}(a)}/d\ln{a}$. It has been shown that in a broad range of cosmologies $f(\Omega_\mathrm{m}(a))\sim \Omega_\mathrm{m}^\gamma(a)$, an empirical relation that we retrieve for the oscillating dark energy cosmologies as well. Deviations with respect to the $\Lambda$CDM expectation however are smaller than for the growth factor itself.\\
The logarithmic derivative of the growth factor can be used in combination with the power spectrum normalization $\sigma_{8}$ to derive the quantity $f(z)\sigma_{8}(z)$, where $\sigma_{8}(z)=\sigma_{8}D_{+}(z)$. It represents the growth rate of structure weighted by a time-dependent normalization. We used the appropriate $\sigma_{8}$ for each model, as reported in Table~\ref{tab:eos_param}. This quantity was recently measured by \cite{Blake2011a} using the WiggleZ Dark Energy Survey data. Measurements were done in four redshift slices using redshift space distortions for the non-linear power spectrum. For more details, we refer to \cite{Blake2011a}. In Figure~\ref{fig:fsigma8}, we compare the analytical predictions for our models with the observational data points. For an easy comparison, we also show the prediction for the fiducial $\Lambda$CDM model. The black filled points represent the WiggleZ measurements with the corresponding error bars.

As noted by \cite{Blake2011a}, the $\Lambda$CDM model prediction of $f(z)\sigma_{8}(z)$ is a very good fit to the WiggleZ data points, but, since the oscillating models here used to have their free parameters calibrated on already existing data, they only slightly differ from the fiducial model. While this statement only had qualitative significance before, we can ground it on a more quantitative basis analysing Figure 8. The WiggleZ data points have a relative accuracy between roughly $9\%$ and $17\%$ up to redshift $z\sim 0.9$ but the oscillating dark energy models differ at most of $8\%$ at $z\sim 0.4$, being therefore all well within the error bars. Also in this case, we can therefore safely assume that at the linear level this class of models is not excluded by the most recent observations.

\subsection{ISW and Rees-Sciama effects}

\begin{figure}
\includegraphics[angle=-90,width=0.45\textwidth]{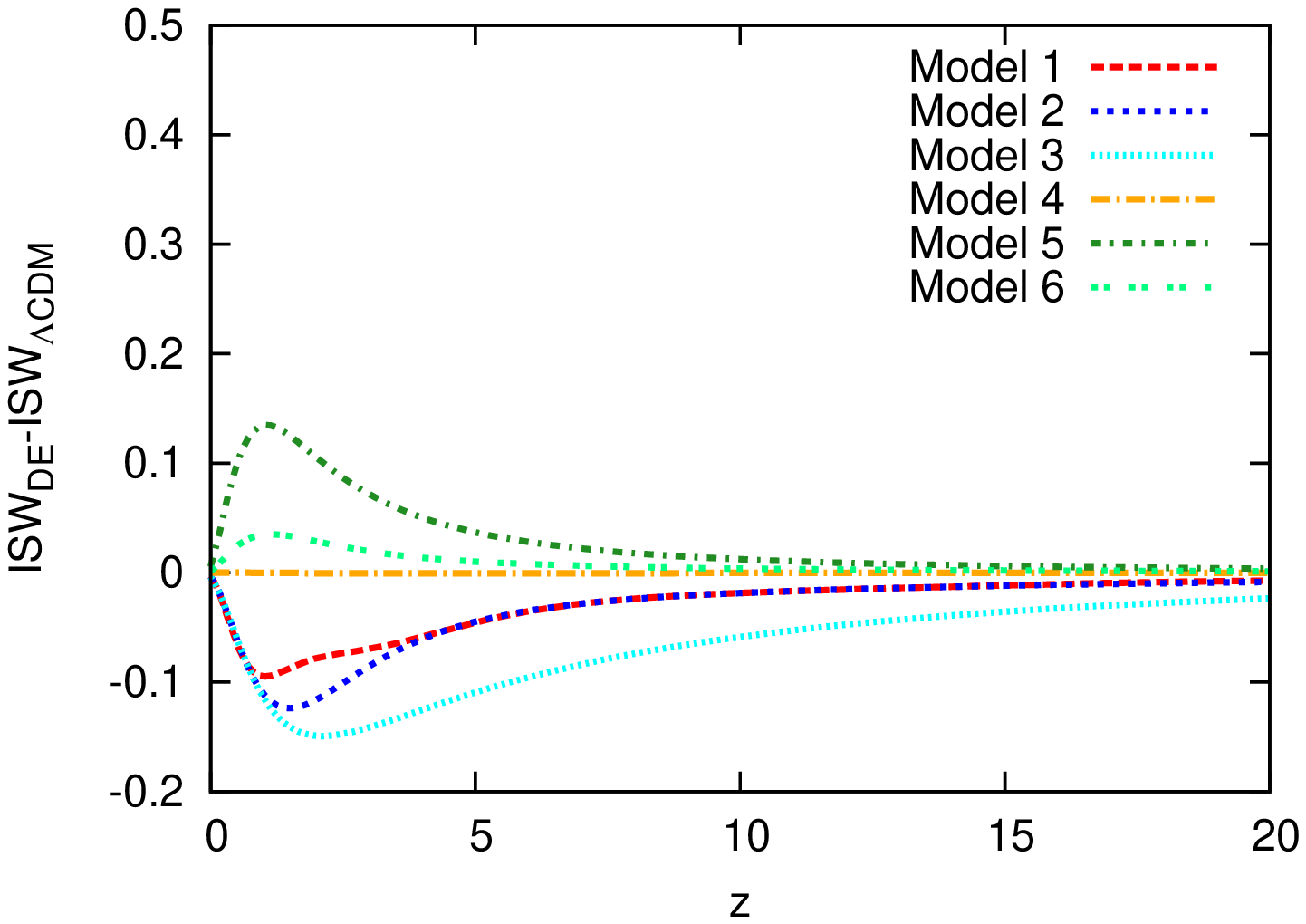}
\includegraphics[angle=-90,width=0.45\textwidth]{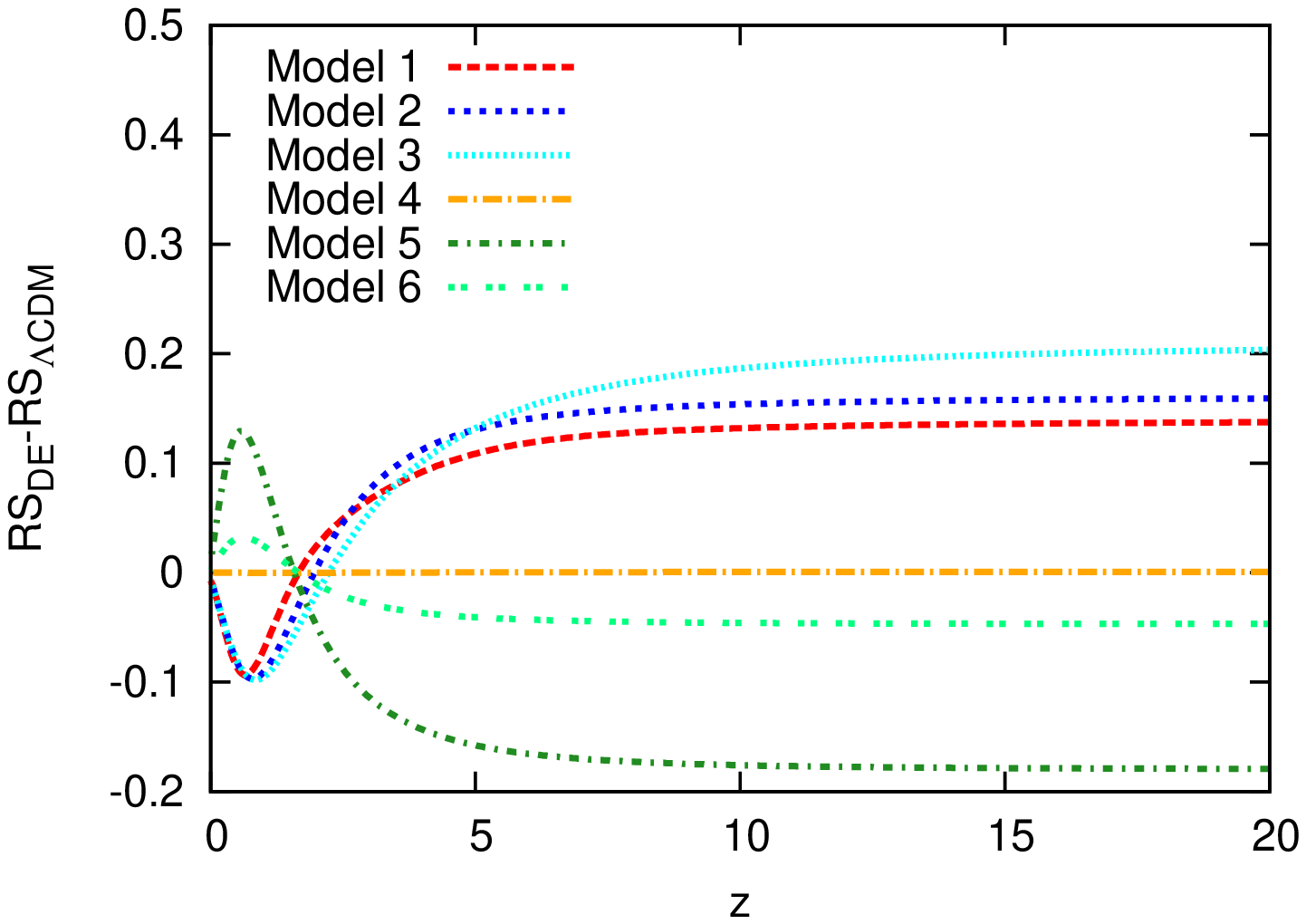}
\caption{Redshift evolution for the quantities describing the ISW (upper panel) and the Rees-Sciama (lower panel) effects. For the ISW effect we plot the function $d\left( D_+(a)/a \right)/da$, while for the RS effect we plot $d\left( D_+^2(a)/a \right)/da$. In both cases, the result for the fiducial model is subtracted from the corresponding quantities for each of the six oscillating dark energy cosmologies. Line types and colours are the same as in Figure~\ref{fig:wz}.}
\label{fig:isw_rs}
\end{figure}

The Sachs-Wolfe (SW) effect \citep{Sachs1967} describes the effect of gravitational potentials on the CMB anisotropy at the last-scattering surface. Photons travelling to an observer encounter variations in the gravitational potential caused by variations in the matter density. Photons climbing out a potential well will be gravitationally redshifted and this will make the region they come from appear colder. Together with this gravitational effect, one has to take into account the time-dilation effect induced by the perturbations: we see the photons as coming from a different spatial hypersurface (labeled by a different scale factor $a(t)$).

The Integrated Sachs-Wolfe (ISW) effect is based on the same principle, only it is given by the gravitational redshift occurring as photons travel through the large scale structure to reach an observer at present time. The ISW effect arises only recently in the cosmic history, as dark energy starts dominating the expansion of the Universe. This means that a non-vanishing ISW effect indicates by itself presence of dark energy if the model is spatially flat, as indeed it is assumed in our case. The ISW effect is sensitive to the derivative of the growth factor, $d\left(D_{+}(a)/a\right)/da$ that vanishes for an EdS universe where $D_{+}(a)=a$. It was detected for the first time by \cite{Boughn2004} using X-ray cluster catalogues. The Rees-Sciama (RS) effect \citep{Rees1968} is very similar to the ISW effect, only it refers to the gravitational redshift induced by non-linear structures only, and hence it is active on much smaller scales. It is mainly sensitive to the function $d\left( D_+^2(a)/a \right)/da$.

In Figure~\ref{fig:isw_rs} we show the difference of the functions probed by the ISW effect (upper panel) and by the RS effect (lower panel) for the six oscillating dark energy models studied here to the same quantities evaluated in the reference $\Lambda$CDM cosmology, as a function of redshift. As can be seen, the ISW effect is preferentially modified at low redshifts, either positively or negatively, except for models 4 where no differences from the $\Lambda$CDM model are seen. At high redshifts deviations with respect to the cosmological constant case tend to disappear, since all the models are very well approximated by an EdS universe. The models showing the largest effect are models 3 and 5, which are the ones having the largest differences in the growth factor. On the other hand, hints of an oscillatory behaviour with redshift are seen for model 1, which has the highest frequency in $w(z)$ among those considered here. All in all, since differences between different models can be quite substantial, high redshift observations could be used in principle to discriminate oscillating quintessence cosmologies. Examples come from cross-correlating galaxies, radio sources or hard X-ray sources and CMB temperature fluctuations, (see \citealt{Fosalba2003,Nolta2004,Scranton2003,Boughn2004,Afshordi2004}). However, one should keep in mind that at high redshift, where differences are more marked, the ISW effect itself tends to disappear.

As for the RS effect, deviations with respect to the cosmological constant expectations are of the same order of magnitude as the ISW effect and differences do not vanish at high redshifts, but reach a somewhat constant value, the exact one depending on the specific model. There is however a very specific redshift pattern according to which the difference with respect to the fiducial $\Lambda$CDM cosmology gets reversed at $z\sim 2$ (except for model 4, that is basically identical to the concordance cosmology). Hence, combining RS effect observations at low and high redshift can improve the discrimination between the models. 

\subsection{The characteristic overdensities $\delta_{\mathrm c}$ and $\Delta_{\mathrm v}$}\label{sect:dcdv}

\begin{figure}
\includegraphics[angle=-90,width=0.45\textwidth]{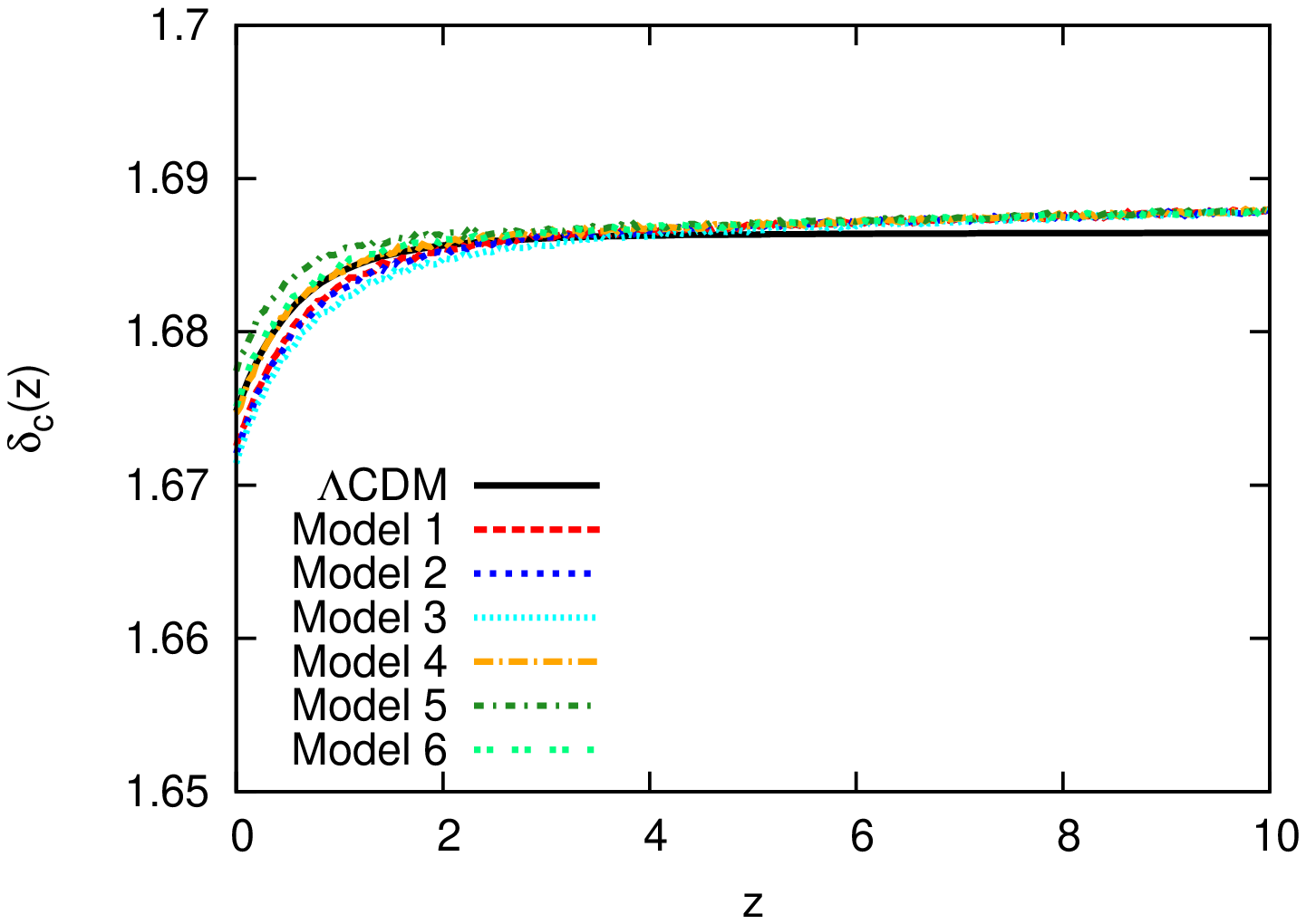}
\includegraphics[angle=-90,width=0.45\textwidth]{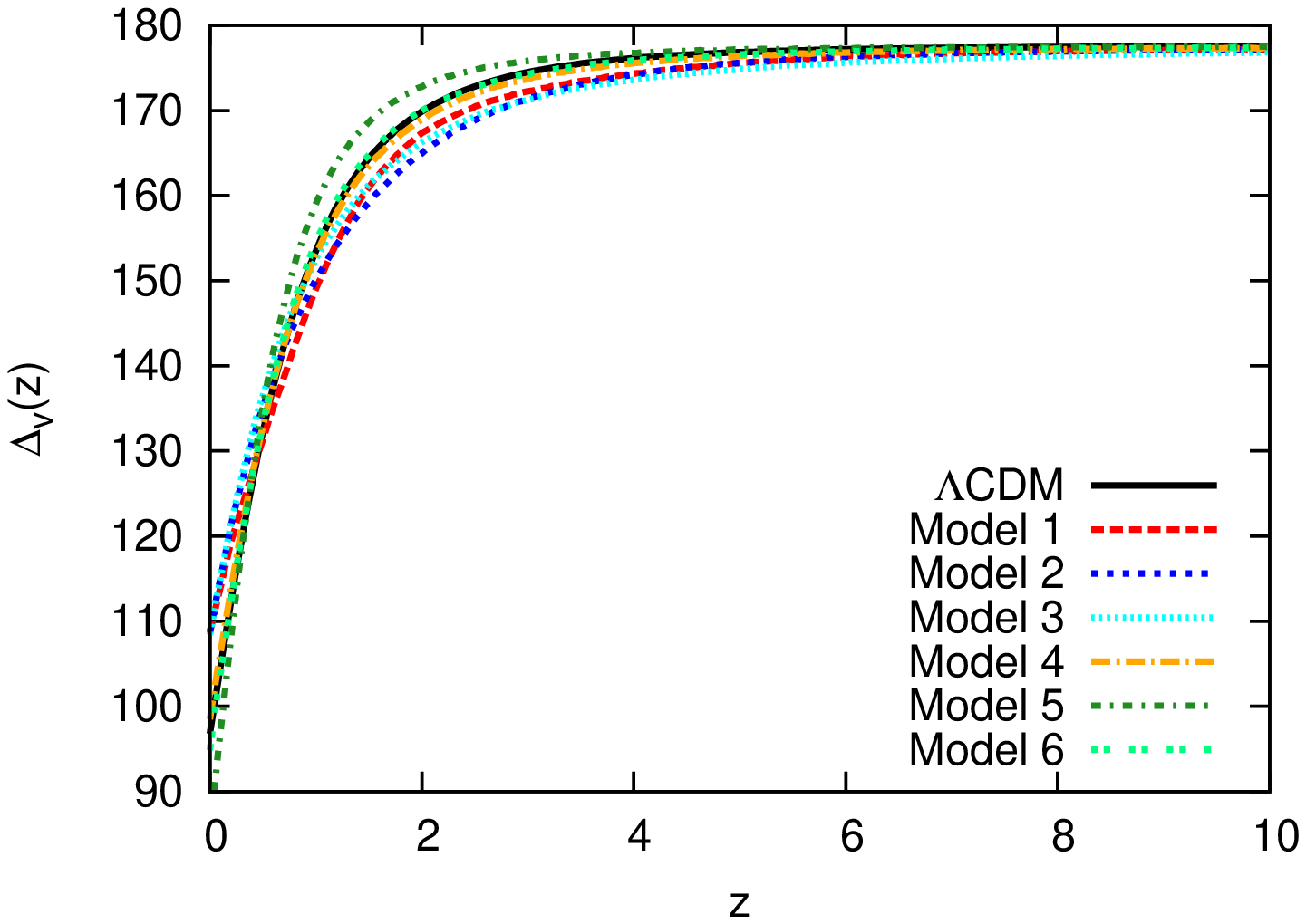}
\caption{The redshift evolution of the linear density contrast parameter $\delta_{\mathrm c}$ (upper panel) and of the virial overdensity parameter $\Delta_{\mathrm v}$ (lower panel) for the six dynamical dark energy models and for  the reference $\Lambda$CDM cosmology. Line types and colours for the different models are as in Figure~\ref{fig:ez_points}.}
\label{fig:spc}
\end{figure}

In this Subsection we present results regarding the time evolution of the linear density contrast parameter for spherical collapse $\delta_{\mathrm c}$ and of the virial overdensity $\Delta_{\mathrm v}$. The main results are reported in Figure~\ref{fig:spc}. In the upper panel we show the time evolution of $\delta_{\mathrm c}$ while in the lower panel we present the time evolution of $\Delta_{\mathrm v}$. Line types and colours are the same as in Figure~\ref{fig:gf}, to which we refer for a detailed explanation. The first thing to note is that, contrary to expectations, the function $\delta_\mathrm{c}(z)$ does not perfectly converge to the EdS value of $\delta_\mathrm{c} \simeq 1.686$ at high redshift. This is a problem of numerical convergence related to the oscillatory nature of $w(z)$, and that is better explored in Appendix~\ref{sect:code}. This fact obviously implies that we should not use this $\delta_\mathrm{c}(z)$ at $z \gtrsim 5-6$. However, all the cosmological tests that we propose in the following that make use of this function are limited to substantially lower redshifts, hence they should be unaffected by this issue.

As mentioned above, differences between the $\delta_\mathrm{c}(z)$ computed in different cosmologies are very mild, being at most of $\sim 1\%$ at $z\lesssim 2$. It is apparently a generic feature of cosmological models displaying a dynamical evolution of the dark energy density that the spherical collapse parameters are only slightly modified with respect to the fiducial $\Lambda$CDM case \citep{Pace2010}. The lower panel of Figure~\ref{fig:spc} shows results for the virial overdensity parameter $\Delta_\mathrm{v}(z)$. In order to evaluate it we used the prescription of \cite{Wang1998}. In this case for $z\gtrsim 6-8$ all the models behave almost exactly as the $\Lambda$CDM cosmology. Differences between different models are of the order of a few percent and are mostly evident at $z\lesssim 3$. We tried to evaluate $\Delta_\mathrm{v}(z)$ adopting a different prescription, detailed in \citet{Wang2006}. As it turns out, differences between different models are very similar to those obtained by using other recipes.

\subsection{Mass function}

An observable quantity depending crucially on the growth factor $D_+(z)$ and on the linear overdensity threshold for collapse $\delta_{\mathrm{c}}(z)$ is the mass function of cosmic structures $n(M,z)$, representing the number of dark matter halos per unit mass and per unit comoving volume present at a certain redshift. Integrals of the mass function over mass can be observed directly by using large cosmological surveys, provided their selection function is well understood. Specifically, the cumulative number density of cosmic structures above a certain limiting mass $M_\mathrm{min}$ (that will depend on the specific survey) at redshift $z$ is simply given by

\begin{figure*}
\includegraphics[angle=-90,width=0.45\textwidth]{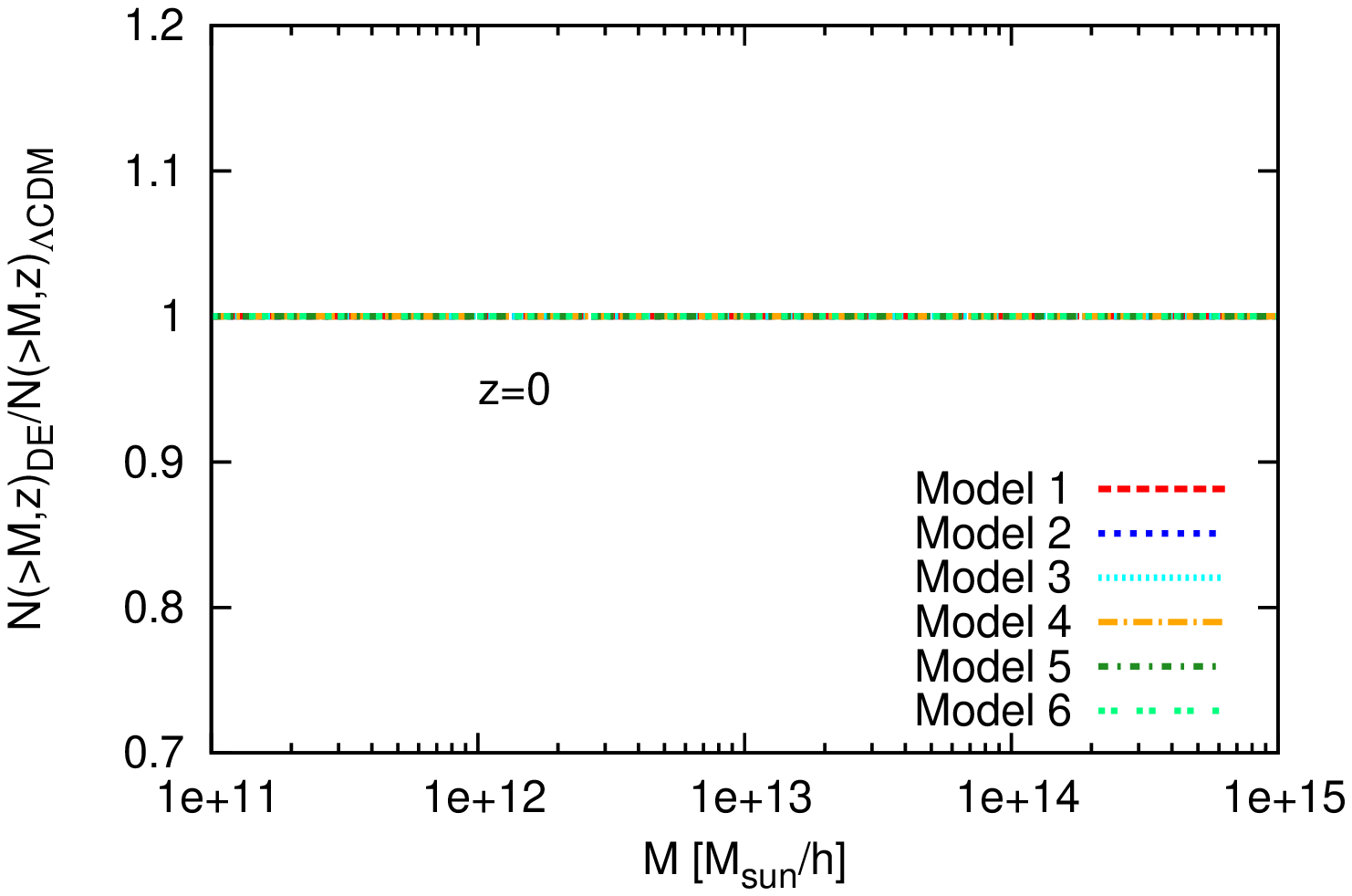}
\includegraphics[angle=-90,width=0.45\textwidth]{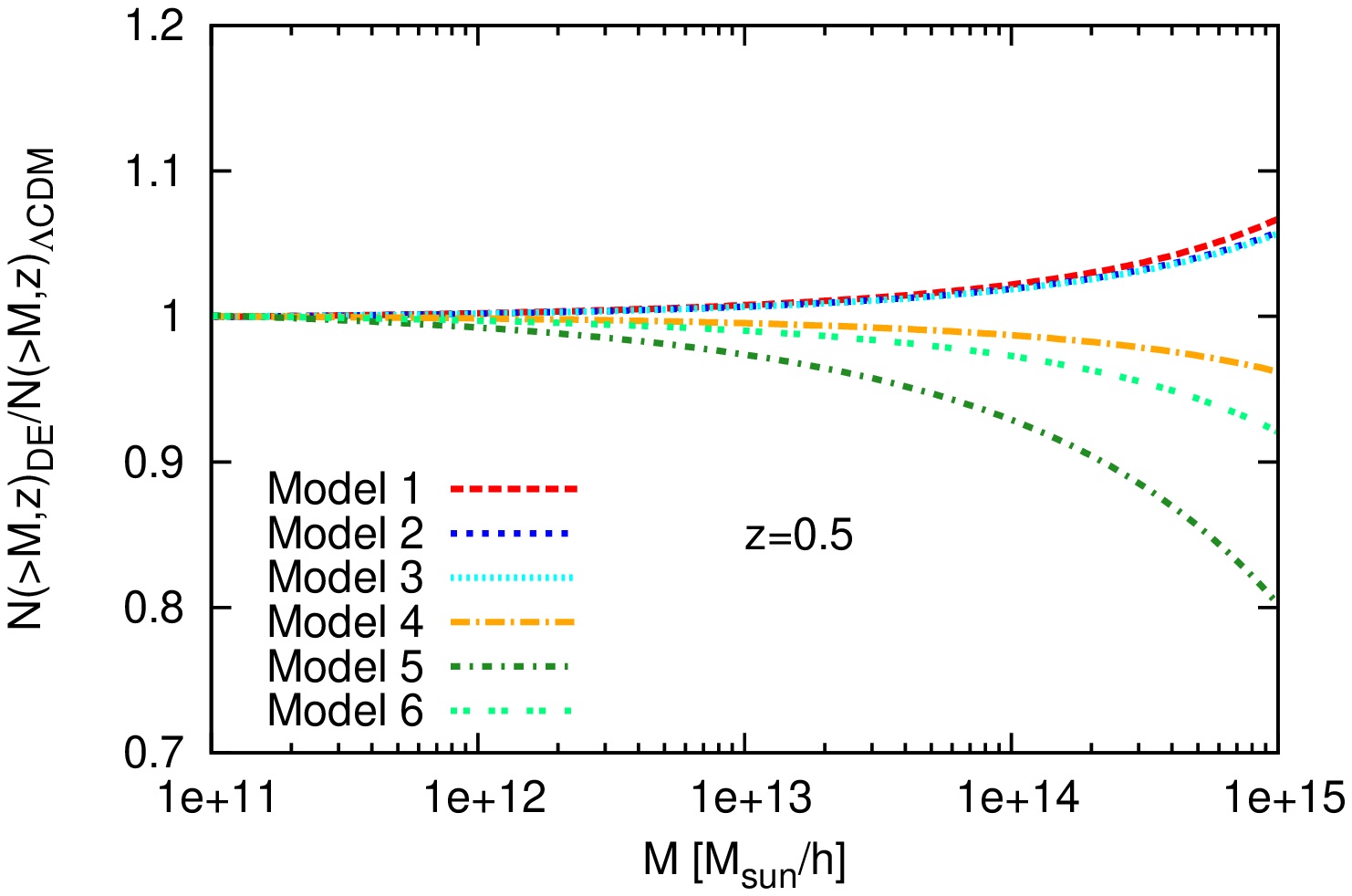}\hfill
\includegraphics[angle=-90,width=0.45\textwidth]{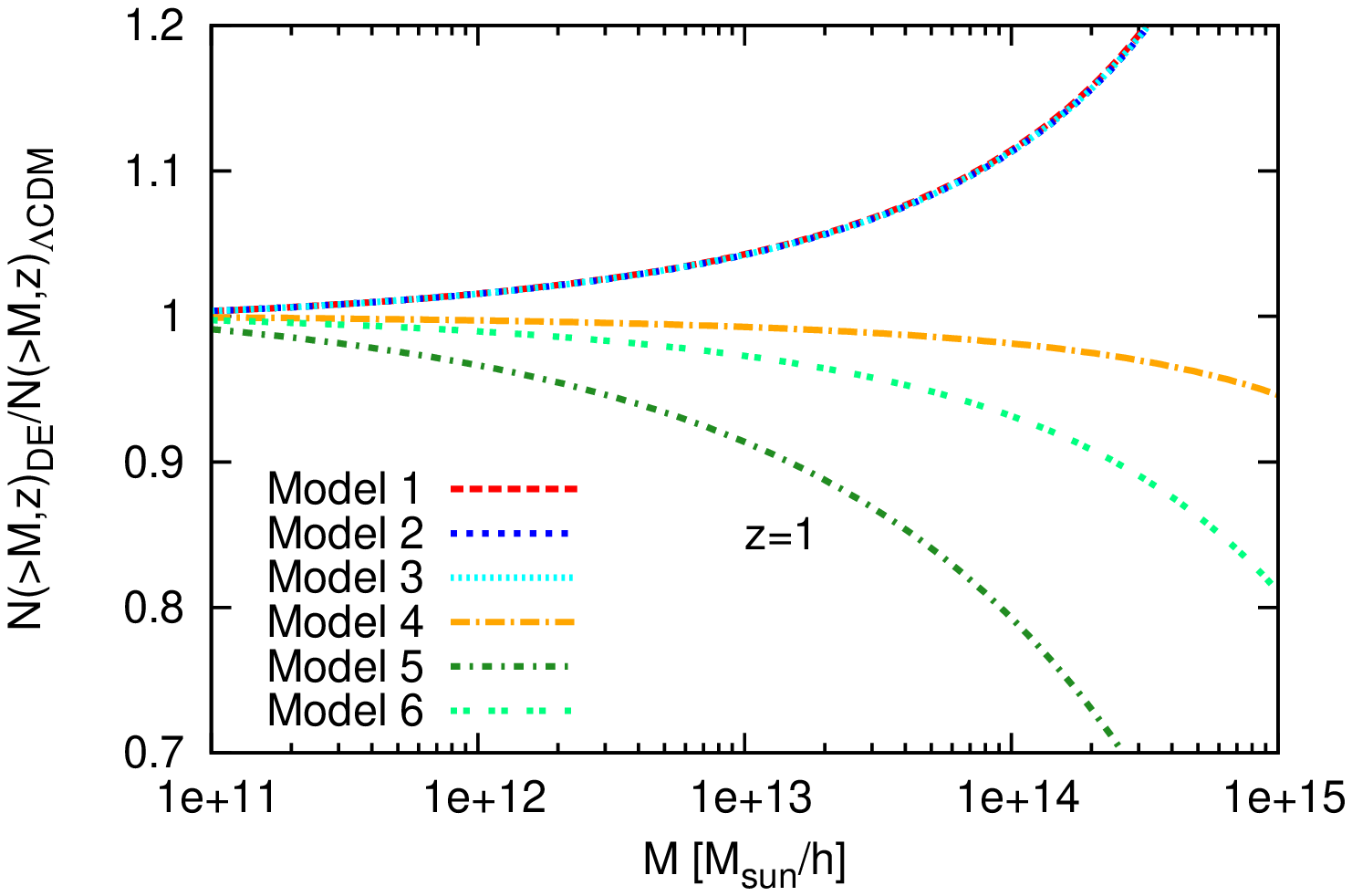}
\includegraphics[angle=-90,width=0.45\textwidth]{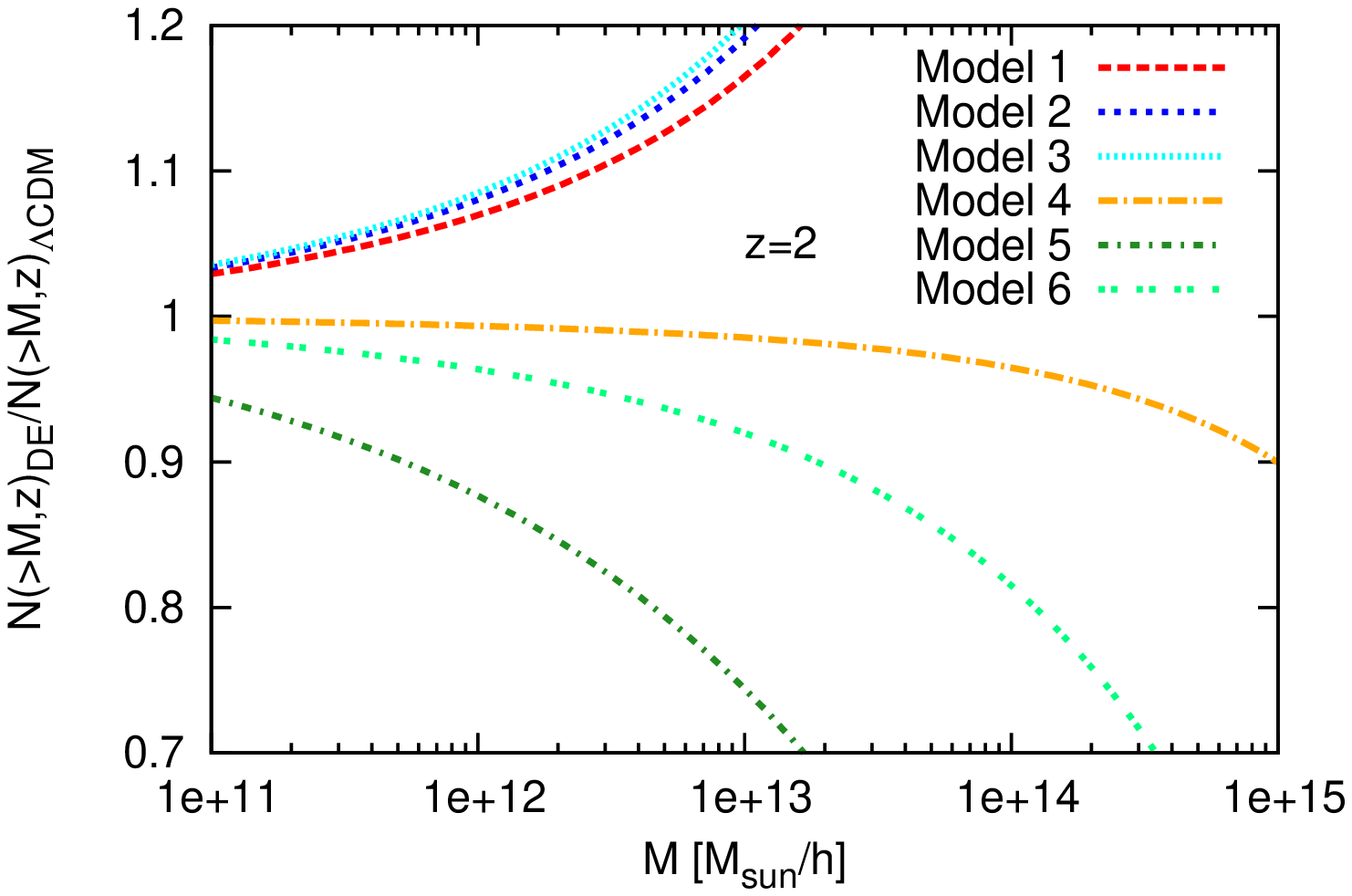}
\caption{Cumulative comoving number density of objects with mass exceeding $M$ at different redshifts. Ratios with respect to $\Lambda$CDM expectations are shown. Selected redshifts are $z=0$ (upper left panel), $z=0.5$ (upper right panel), $z=1$ (bottom left panel) and $z=2$ (bottom right panel). Line styles and colours for the different models are as in Figure~\ref{fig:wz}. The upper left panel is unity by normalization, as explained in the text.}
\label{fig:mf}
\end{figure*}

\begin{equation}
N(>M_\mathrm{min},z)=\int_{M_{\mathrm{min}}}^{\infty}\mathrm{d}M n(M,z)~.
\end{equation}

The mass variance is another key ingredient for the mass function formalism, and is identified by

\begin{equation}
\sigma^{2}_{M}=\frac{1}{2\pi^{2}}\int_0^{+\infty}k^2T^2(k)W^2_R(k)P_0(k)dk\;.
\label{eqn:sigma}
\end{equation}
In Eq.~(\ref{eqn:sigma}) $P_0(k)$ represents the primordial matter power spectrum, $T(k)$ is the matter transfer function, while $W_R(k)$ is the Fourier transform of the real space top-hat window function. Since the only difference induced by oscillating dark energy in the primordial matter power spectrum is given by the different normalization $\sigma_8$, that as we shown above is very slight, we expect only minor differences in the mass variance as well. In this work, to evaluate the mass function, we used the Sheth-Tormen expression \citep{Sheth1999}.

In Figure~\ref{fig:mf} we show the cumulative mass function for the six oscillating quintessence cosmologies considered in this work at different redshifts, divided by the corresponding quantity evaluated for the reference $\Lambda$CDM model. From this Figure we note first of all that there are no differences between different models at $z = 0$. This is due to our choice of the normalization, namely that the ratio $\delta_\mathrm{c}(z=0)/\sigma_8$ should be conserved for all models. Significant differences start instead to appear at higher redshifts, where we notice that the six models can be broadly divided into two groups. Models 1, 2, and 3 show more objects with respect to the $\Lambda$CDM case, while models 4, 5, and 6 show less objects. These differences are consistent with the latter group having a lower growth factor with respect to the former and a higher critical linear overdensity $\delta_\mathrm{c}(z)$, which makes more difficult for density perturbations to collapse into bound structures. As one could naively expect, deviations increase with increasing redshift and are most notable in the very high mass tail, since rare events are very sensitive even to small fluctuations in the expansion history. As an example, differences in the abundance of massive galaxy clusters $M\gtrsim 5\times 10^{14} M_\odot/h$ range from $\sim 5-10\%$ at $z = 0.5$ up to $\sim 30\%$ at $z = 1$. Large galaxy groups that can be expected to be found at $z = 2$ ($M \gtrsim 5\times 10^{13} M_\odot/h$) are $\sim 20-30\%$ more abundant in models belonging to the first group, and up to $\sim 40-50\%$ less frequent in models of the second group. We also note that models showing an enhancement in the abundance of cosmic structures are those having more dark energy at high redshift.

It should be noted that up to now we only considered the comoving number density of objects, that is we did not take into account possible effects deriving from changes in the cosmic volume induced by the presence of oscillating dark energy. We will include this additional ingredient shortly. For the time being, we can conclude that the impact of oscillating quintessence on the counts of cosmic structure can be quite substantial, especially at high redshift, thus implying that a detection might be possible with future large cluster surveys. 

In order to establish a more direct link with observations, we forecast the redshift distribution of galaxy clusters that, in each of the cosmological models considered in this work, will be observed by upcoming wide field surveys with cluster selection based both on Xray and weak-lensing data. In order to do that we need to define a redshift dependent minimum mass for the observed objects, and integrate the mass function above that threshold. We assume that precise estimates for the masses of these objects will be available, which is realistically expected if a robust multiwavelength follow-up will be performed. 

The first survey that we consider is a wide field X-ray survey on the model of the upcoming \emph{e}ROSITA\footnote{http://www.mpe.mpg.de/erosita/} one. In order to determine the minimum mass of clusters that will compose the X-ray catalogue we need a scaling relation between the observable at hand (in this case the X-ray flux) and the true mass. First of all, knowing the redshift of the cluster, the measured bolometric flux can be related to the intrinsic bolometric luminosity as $L(M,z) = 4\pi f(M,z) d_\mathrm{L}^{2}(z)$, where $d_\mathrm{L}(z)$ is the luminosity distance (see upper panel of Figure \ref{fig:dist_time}). In reality the X-ray bolometric flux is almost never measured, rather the X-ray photon counts in a certain energy band are used. For the specific case of an \emph{e}ROSITA-like X-ray survey, we adopted the band $[0.5,2.0]$ keV, where the threshold flux is expected to be $f_\mathrm{min} = 3.3\times 10^{-14}$ erg$/($s cm$^2)$. In order to estimate the band flux for a cluster of a given mass placed at a given redshift we modelled the related intra-cluster medium using a Raymond-Smith plasma model \citep{Raymond1977} as implemented in the \emph{XSPEC} software package \citep{Arnaud1996}, with metal abundance $Z=0.3 Z_{\odot}$ \citep{Fukazawa1998,Schindler1999}. The plasma model has been normalized so as to reproduce the bolometric luminosity expected from the scaling relation adopted by \citet*{Fedeli2009}, namely

\begin{equation}
L(M,z)=1.097\times 10^{45}~\mathrm{erg/s}~(ME(z))^{1.554}~,
\end{equation}
where the mass must be inserted in units of $10^{15}M_\odot/h$. This relation results from the combination of two scaling laws, one relating the mass to the X-ray temperature, and the other relating the temperature to the luminosity. See \citet*{Fedeli2009} for additional details.

The second case that we consider is representative of cluster catalogues selected through their weak-lensing signal. Massive galaxy clusters can be selected as high S/N peaks in the weak lensing map produced by weak-lensing surveys. The S/N strength will also be used as a proxy for their mass, although a robust multiwavelength follow-up program will be necessary in order to have more precise estimates. For the minimum cluster mass entering this catalogue, we adopted the calculations of \citet*{Berge2010}. In particular, we refer to their Figure~1, where they present the selection function for a \emph{Euclid}-like survey\footnote{http://sci.esa.int/science-e/www/area/index.cfm?fareaid=102} \citep{Laureijs2009,Laureijs2011} in the mass-redshift plane, assuming a number density of background galaxies of $\bar n_{\mathrm{g}}=40$ arcmin$^{-2}$. We considered the contour referring to a S/N threshold of $5$, threshold that was shown to be a good choice to minimize spurious detections in the weak lensing maps \citep[see][]{Pace2007.1}.

\begin{figure}
\includegraphics[angle=-90,width=0.45\textwidth]{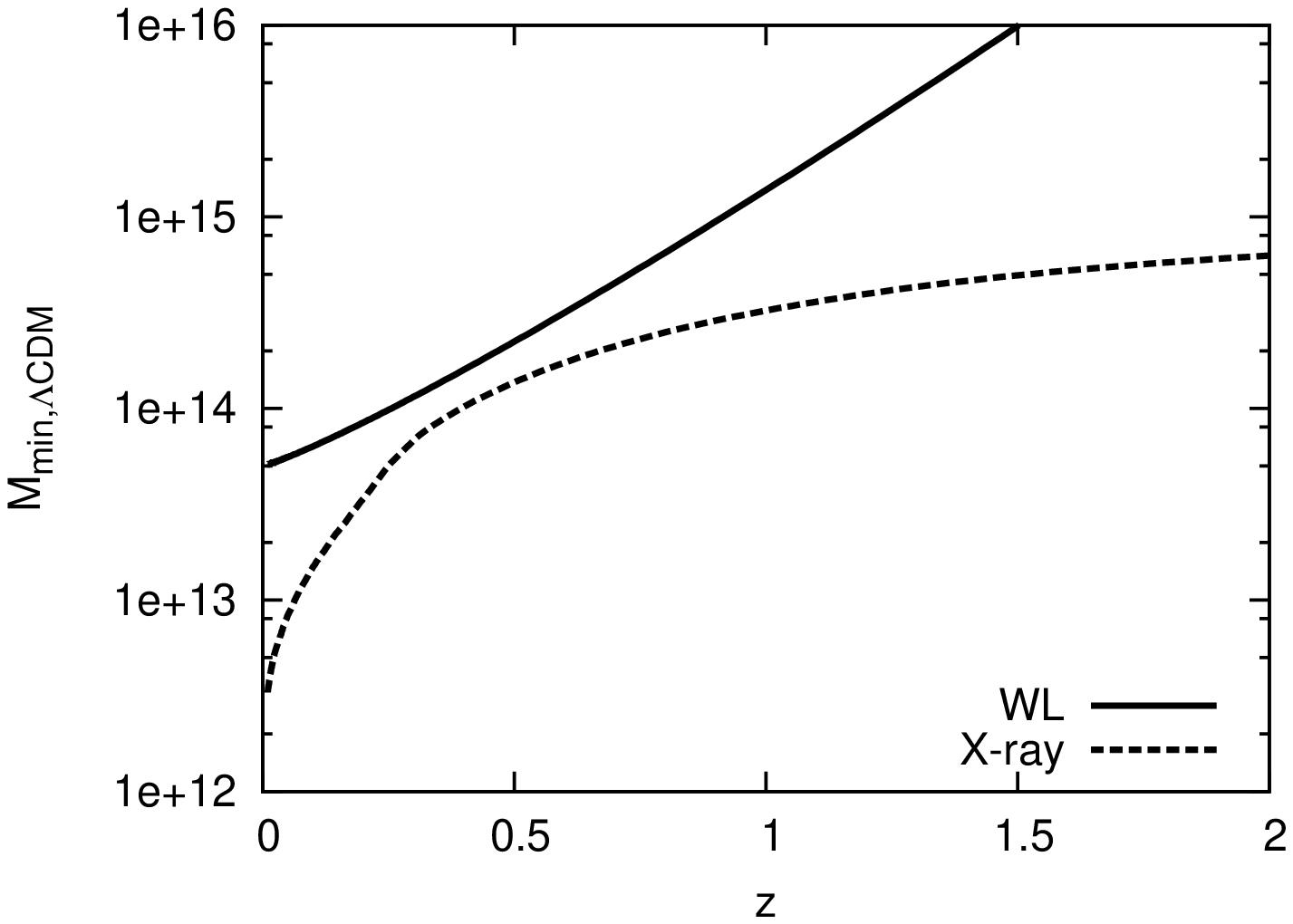}
\includegraphics[angle=-90,width=0.45\textwidth]{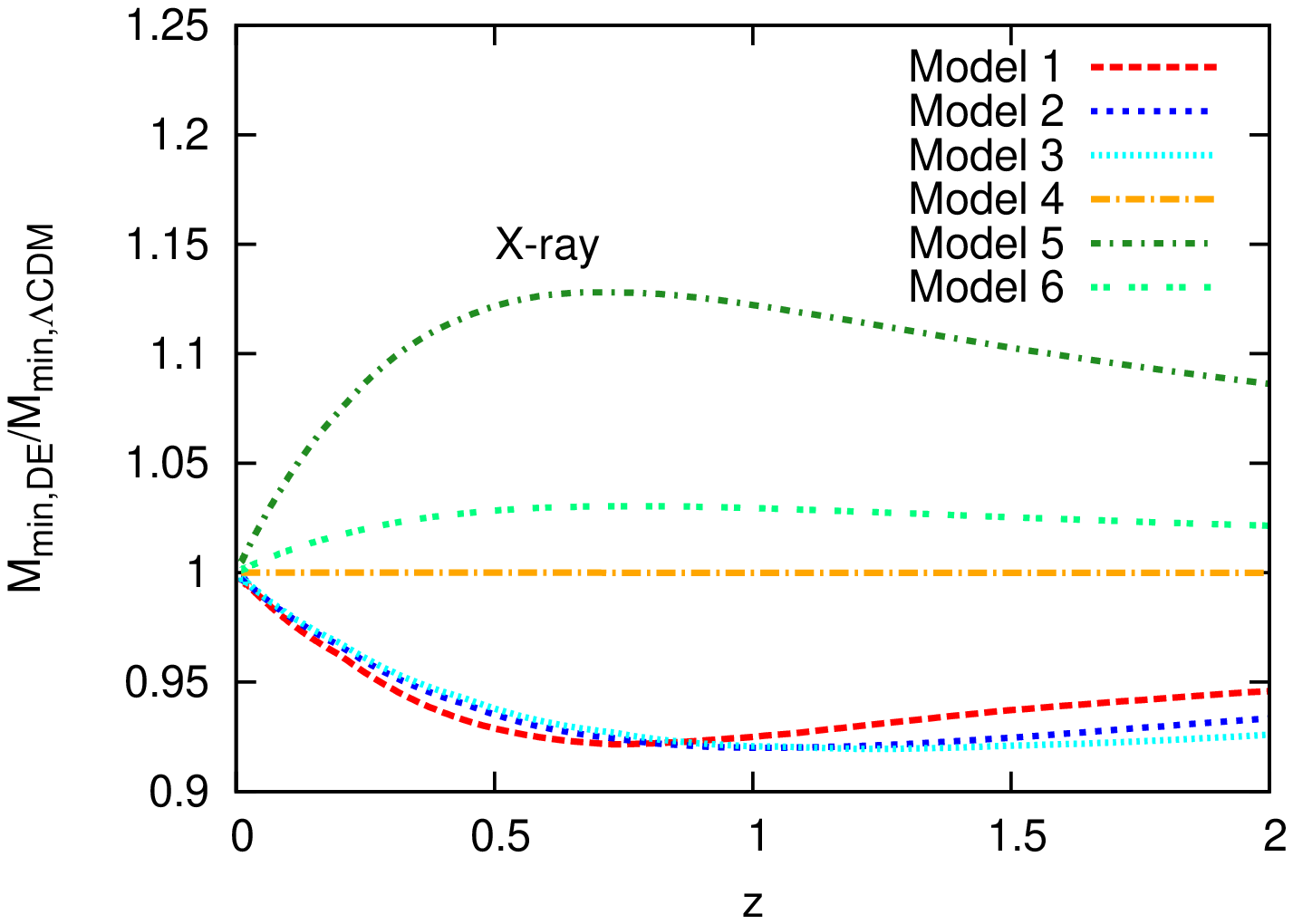}
\caption{emph{Upper panel}. Minimum mass for weak-lensing survey (solid line) and X-ray survey (dashed line) for the reference $\Lambda$CDM model. \emph{Lower panel.} Minimum mass for the \emph{e}ROSITA cluster catalogue as a function of redshift, presented as ratios between the oscillating dark energy models and the $\Lambda$CDM cosmology estimates.}
\label{fig:min_mass}
\end{figure}

In Figure~\ref{fig:min_mass} we present our results for the minimum mass of the catalogues. In the upper panel we show the minimum mass for the adopted weak-lensing survey (solid line) and X-ray survey (dashed line) for the $\Lambda$CDM model. We see that as expected both minimum masses increase with redshift in order to have the same flux or S/N ratio. The X-ray mass increases from few times $10^{13}M_\odot/h$ at $z\simeq 0$ till $\simeq 6\times 10^{14}M_\odot/h$ at $z=2$. As evident, since the flux limit is constant in redshift, the redshift dependence of the mass can be very well approximated by a parabola, therefore compensating the increase of the luminosity distance (entering quadratically in the relation between flux and luminosity). The minimum mass for a weak-lensing survey is increasing much faster with redshift since the lensing efficiency drops very fast to zero if the lens approaches the sources. \\
In the lower panel we show the minimum cluster mass for the X-ray catalogue in each dark energy cosmology considered here, divided by the same quantity estimated for the reference $\Lambda$CDM model (lower panel). We see that differences in the minimum observed mass are strongly related to the differences in the Hubble function and in the luminosity distance, as one might naively have expected. Specifically, since the minimum mass depends on the square of the luminosity distance, even small variations of the latter turn out to affect the former at the level of $\sim 10\%$.

\begin{figure}
\includegraphics[angle=-90,width=0.45\textwidth]{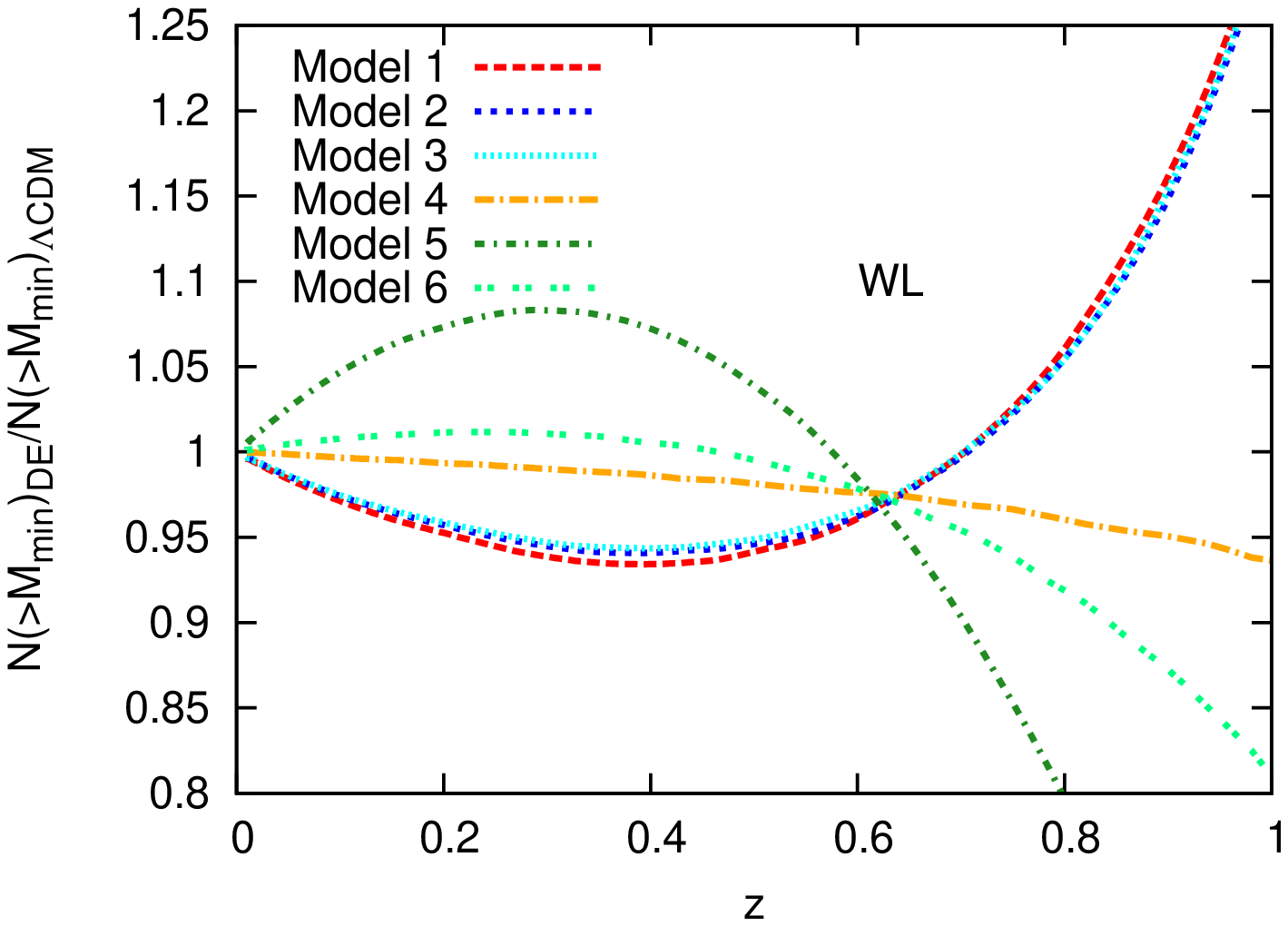}
\includegraphics[angle=-90,width=0.45\textwidth]{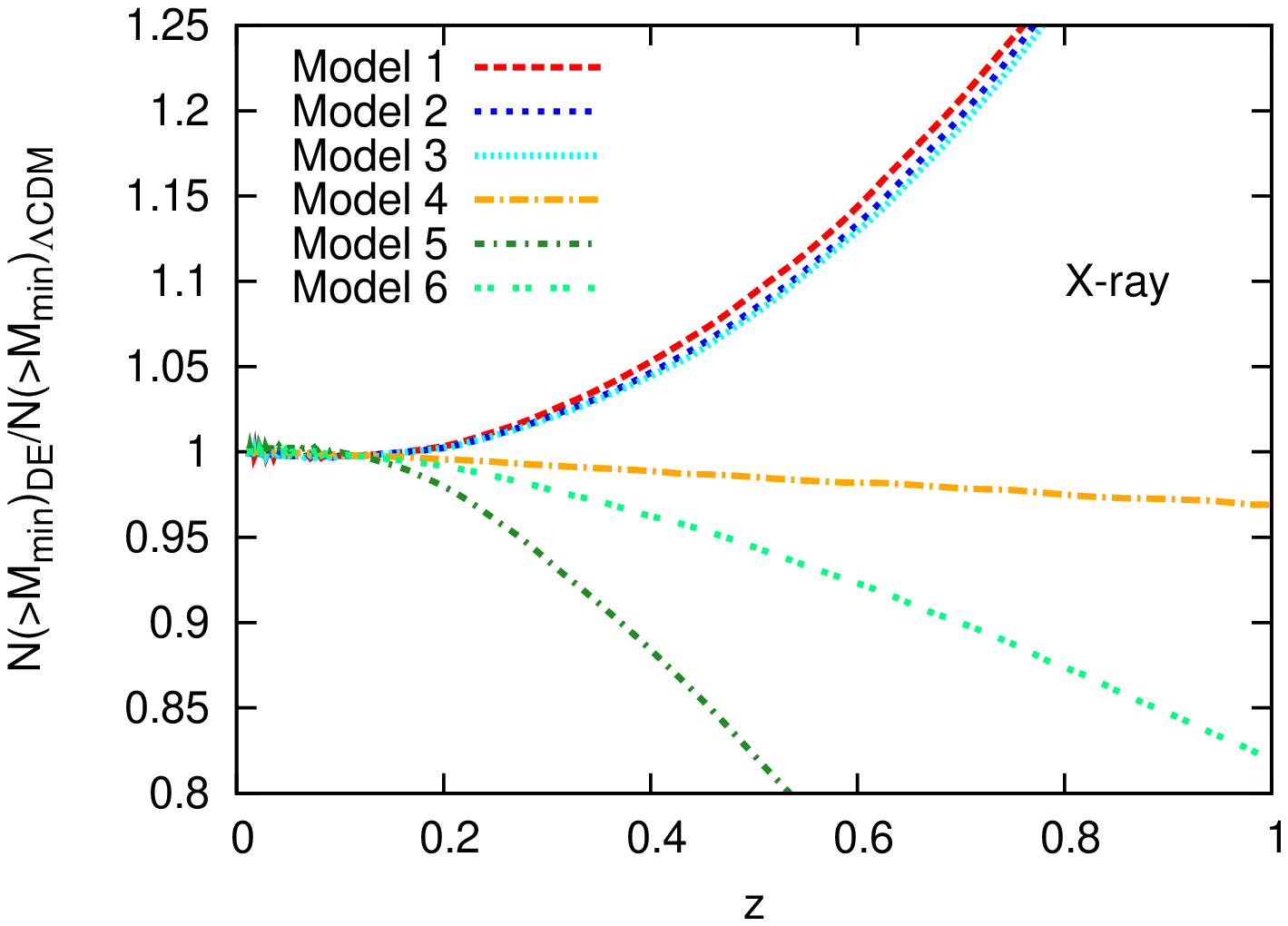}
\caption{Cluster redshift distribution, as a function of redshift, presented as ratios between the oscillating dark energy models and the $\Lambda$CDM cosmology estimates. \emph{Upper panel.} Number of objects above the minimum mass for a \emph{Euclid}-like survey. \emph{Lower panel.} Number of objects above the minimum mass for an \emph{e}ROSITA-like survey. Line styles and colours for the different models are as in Figure~\ref{fig:wz}.}
\label{fig:mnz}
\end{figure}

In Figure~\ref{fig:mnz} we present the redshift distributions expected for the weak-lensing survey (upper panel) and the X-ray survey (lower panel) we considered, respectively. The redshift distribution is defined as

\begin{equation}
\mathcal{N}(z)=\frac{dV(z)}{dz}~N(>M_\mathrm{min}(z),z),
\end{equation}
where $M_\mathrm{min}(z)$ is the minimum observed mass for the survey at hand and $dV(z)/dz$ is the comoving volume element contained in the unit redshift. The first thing to note is that the deviations in the redshift distributions induced by the presence of oscillating dark energy are quite substantially different for an X-ray survey and and a cosmic shear survey. This is likely due to the fact that these surveys have remarkably different selection functions, that sample quite distinct regions of the mass-redshift plane, where the impact of oscillating dark energy is necessarily different. Let us consider first the X-ray \emph{e}ROSITA-like survey. In this case the redshift distributions computed for different cosmologies are almost identical at very low redshift, while substantial deviations are visible at higher redshift. Specifically, the abundance of clusters is incremented by up to $\sim 20\%$ at $z\sim 0.8$ for models 1, 2, and 3, and decreased by the same amount or more for the other models.

For the weak lensing survey the situation is totally different. Cosmological models differ from each other by up to $\sim 5-10\%$ already at relatively low redshift. Deviations from the $\Lambda$CDM curiously vanish at $z\sim 0.6$ for all models, and then grow again, but with opposite sign, for higher redshifts. In addition to the different selection function, one additional difference between the X-ray survey and the cosmic shear survey considered here is that in the former case the minimum mass depend on cosmology, while in the latter case it is model independent since what we measure is directly related to the mass of the cluster.

\subsection{Dark matter power spectrum}

An important tool that can be used to infer the statistical properties of a cosmological model is the fully non-linear matter power spectrum. Observationally, this can be estimated both by using the distribution of tracers such as galaxies and galaxy clusters (provided their non-linear bias is understood) and through cosmic shear (that however returns only a projected version of the three-dimensional spectrum). The matter power spectrum can be studied in the non-linear regime by means of numerical N-body simulations or semi-analytic prescriptions that are fitted against them \citep{Peacock1996,Smith2003}. The shortcoming of such fitting formulas stays in the fact that they have limited tests of validity, usually restricted to the concordance $\Lambda$CDM cosmology and scales $k \lesssim 50-100~h/$Mpc at $z = 0$.

An alternative approach, that we exploited, is based on the halo model developed by \cite{Ma2000a} and \cite{Seljak2000} (see \citealt{Cooray2002} for a comprehensive review). The halo model has a physical motivation, and relies on ingredients, such as the average dark matter halo density profile and mass function, whose universality is much better established than for the non-linear matter power spectrum. Additionally, using this formalism the matter power spectrum calculation can be pushed to very small scales, in principle as small as the structure of cold dark matter halos has been studied. Within the halo model the full power spectrum is given by the sum of two terms: one (the $2-$halo term) dominates on large scales and it depends on the correlation of individual halos; the other (the $1-$halo term) dominates on small scales and it is sensitive to the inner structure of the halos. The ingredients needed in order to apply the halo model formalism are the halo mass function, the halo linear bias, and the halo mass density profile.

\begin{figure}
\includegraphics[angle=-90,width=0.45\textwidth]{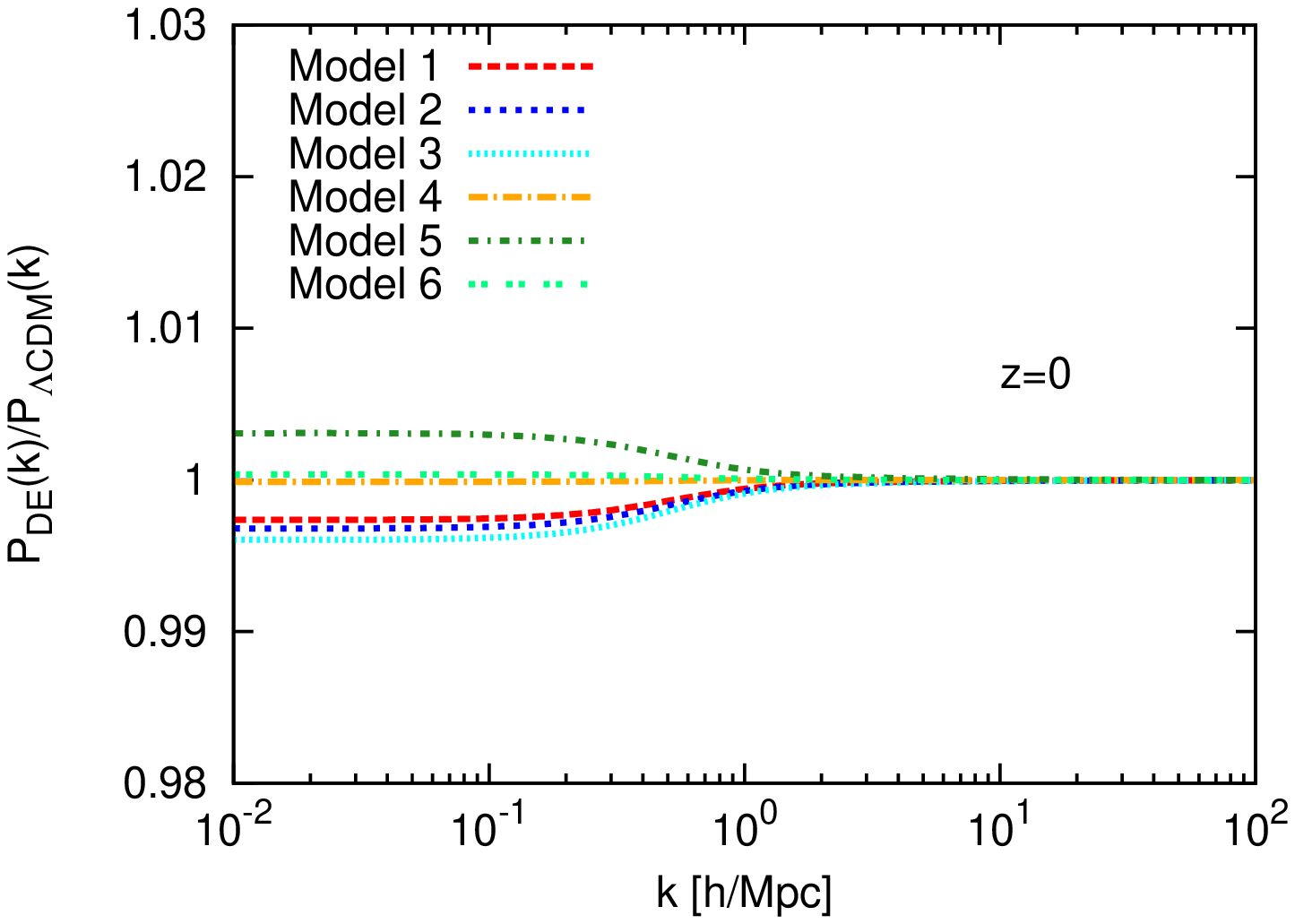}\hfill
\includegraphics[angle=-90,width=0.45\textwidth]{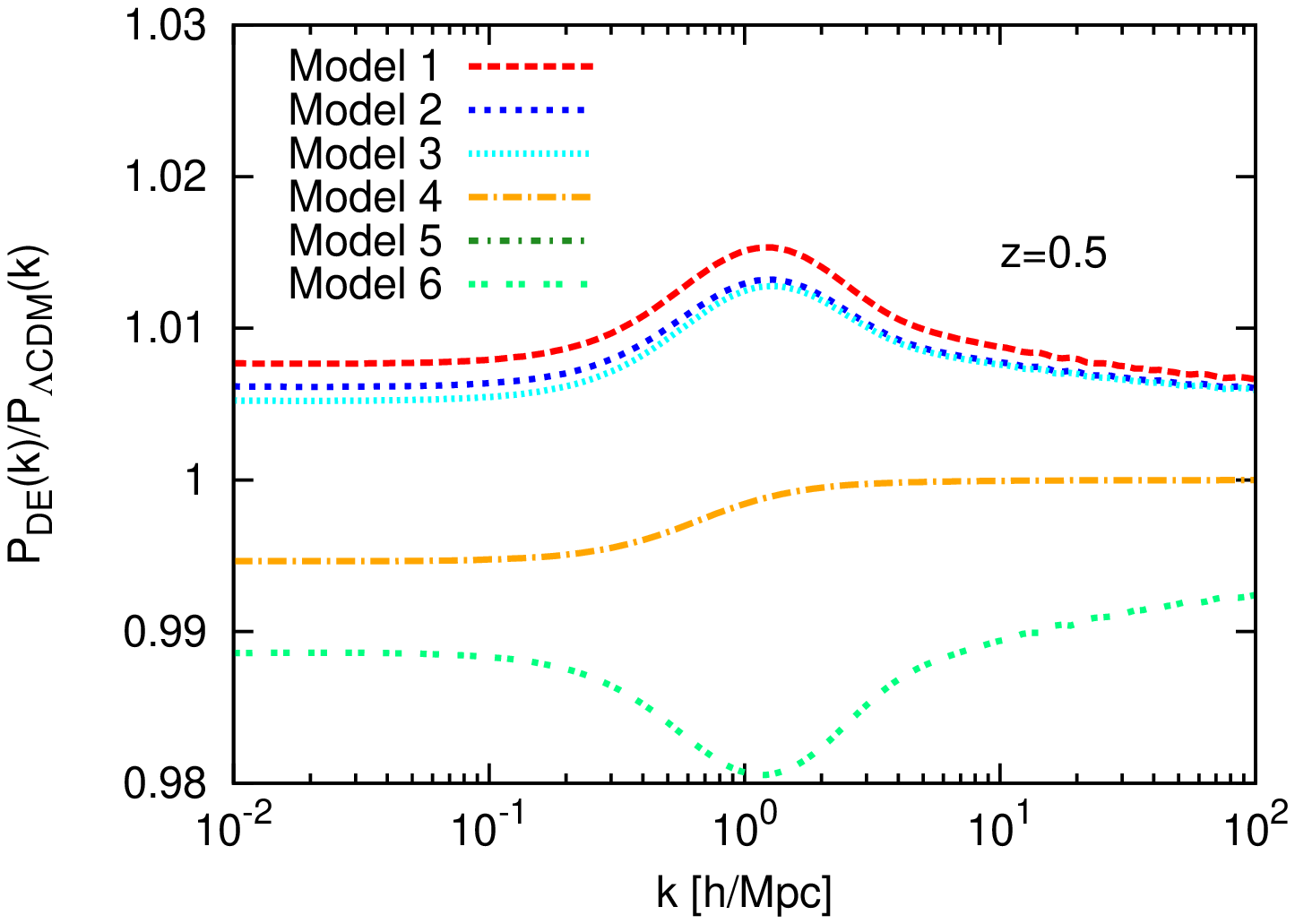}
\caption{The fully non-linear dark matter power spectrum at $z=0$ (upper panel) and $z=0.5$ (lower panel) according to the halo model prescription. Ratios with respect to the $\Lambda$CDM expectation are shown. Line types and colours for the different models are as in Figure~\ref{fig:wz}.}
\label{fig:DMps}
\end{figure}

The behaviour of the mass function in the various oscillating dark energy cosmologies considered here has been described in the previous Subsection. For the average internal structure of dark matter halos we adopt the \citet*{Navarro1997} (NFW henceforth) density profile,

\begin{equation}
\rho(r)=\frac{\rho_{\mathrm{s}}}{(r/r_{\mathrm{s}})(1+r/r_{\mathrm{s}})^{2}}~,
\end{equation}
where $\rho_\mathrm{s}$ is a density scale while $r_\mathrm{s}$ is the radius at which the logarithmic slope of the profile equals $-2$. The parameter $r_\mathrm{s}$ is related to the virial radius $R_\mathrm{v}$ of the structure by the concentration, $c\equiv R_\mathrm{v}/r_\mathrm{s}$. The concentration is in turn depending on the total mass $M$ of the object, in that dynamically younger structures have on average larger masses and lower concentrations. We adopt the following concentration-mass relation,

\begin{equation}\label{eqn:conc}
c(M,z)=\frac{10}{1+z}\left[\frac{M}{M_{*}(z)}\right]^{-0.15}~,
\end{equation}
inspired by \citet{Huffenberger2003} and tested to give good agreement with fits to N-body simulations in the framework of the $\Lambda$CDM cosmology. In the previous equation $M_*(z)$ represents the characteristic collapsing mass at a given redshift, defined implicitly as $D_+(z)\sigma_{M_*} = \delta_\mathrm{c}(z)$.

As for the large-scale bias, which is needed in the $2-$halo part of the power spectrum, we adopt the prescription by \citet*{Sheth2001},
\begin{eqnarray}
b(M,z)&=&1+a\frac{\delta_{\mathrm{c}}(z)}{D^{2}_{+}(z)\sigma^{2}_{M}}-\frac{1}{\delta_{\mathrm{c}}(z)}+
\nonumber\\
&+&\frac{2p}{\delta_{\mathrm{c}}(z)}
\left[\frac{(D_{+}(z)\sigma_{M})^{2p}}{(D_{+}(z)\sigma_{M})^{2p}+(\sqrt{a}\delta_{\mathrm{c}}(z))^{2p}}\right]~,
\end{eqnarray}
where $a=0.75$ and $p=0.3$.

In Figure~\ref{fig:DMps} we show the ratio of the matter power spectrum computed in each of the six oscillating dark energy models explored here to the same function in the concordance $\Lambda$CDM cosmology, as a function of wavenumber. In the upper panel we plot the ratio at $z=0$ while in the lower panel we consider the ratio at $z=0.5$. It is immediately evident that at the present time all cosmologies tend to have the same non-linear matter power spectrum at small scales. This is due to the fact that in the concentration-mass relation that we adopted (Eq.~\ref{eqn:conc}) all the cosmology dependence is encapsulated in the characteristics non-linear mass $M_*(z)$. However, due to the fact that $D_+(z=0)=1$, and $\delta_\mathrm{c}(z=0)/\sigma_8$ is the same for all cosmological models, the concentration for a given mass at $z=0$ is also unchanged. Hence, since dark matter halos share always the same inner structure, they produce an identical dark matter correlation function at small scales. The situation changes at $z>0$. In fact, as we can see from the lower panel of Figure~\ref{fig:DMps}, the matter power spectra now differ from the $\Lambda$CDM case at all scales. Differences are however at the level of $\sim 1\%$ at most and model 4 in particular does not show strong differences from the reference model. One might argue that the inner structure of dark matter halos should change at least a bit due to oscillatory dark energy, even at $z=0$. Since however we have no indication on how this is expected to happen, we stick to our original choice. At very large scales the power spectrum approaches the linear primordial one, hence the deviations with respect to the fiducial cosmology are induced only by differences in the normalization $\sigma_8$ (squared), that are always below $1\%$.

\subsection{Cosmic shear}\label{sect:shear}

As mentioned above, the power spectrum of cosmological weak lensing is a projected (and weighted) version of the fully non-linear matter power spectrum. What is actually measured is the effective convergence power spectrum (equivalent to the shear and reduced shear power spectra), given by

\begin{equation}\label{eqn:wlps}
P_\kappa(\ell)=\frac{9H^{4}_{0}\Omega^{2}_{\mathrm{m},0}}{4}\int_{0}^{\chi_{\mathrm{H}}}P\left[\frac{\ell}{f_{K}(\chi)},\chi\right]\frac{W^{2}(\chi)}{a^{2}(\chi)}d\chi\;, 
\end{equation}
where $a(\chi)$ is the scale factor, $\chi$ the comoving distance up to scale factor $a$, and $f_{K}(\chi)$ the comoving-angular diameter distance which depends on $K$, the spatial curvature of the Universe. The integral in the previous equation formally extends up to the horizon size $\chi_{\mathrm{H}}$, however since the number density of sources (see below) drops to zero much before that, the integral can be effectively truncated at $z \sim 10$. 

\begin{figure}
\includegraphics[angle=-90,width=0.45\textwidth]{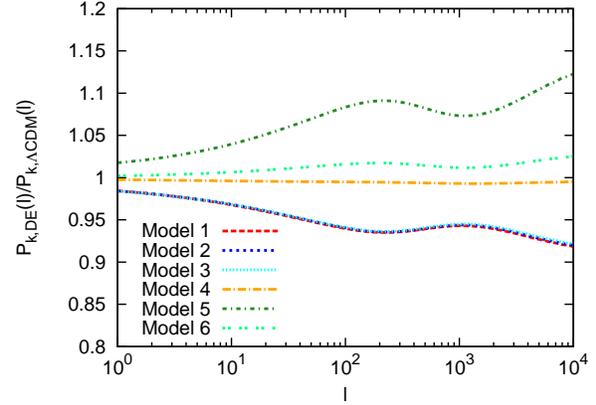}
\caption{The effective convergence power spectrum for each of the six oscillating dark energy cosmologies considered in this work. We show ratios with respect to the $\Lambda$CDM power spectrum. Line styles and colours for the different models are as in Figure~\ref{fig:wz}.}
\label{fig:WLps}
\end{figure}

The function $W(\chi)$ is an integration kernel depending on $n(z(\chi))$, the redshift distribution of background sources. The kernel can be written as

\begin{equation}
W(\chi)=\int_{\chi}^{\chi_{\mathrm{H}}}n(\chi^{\prime})\frac{f_K(\chi-\chi^{\prime})}{f_K(\chi^{\prime})}d\chi^{\prime}\;.
\end{equation}
The functional form for the redshift distribution of the sources has to be inferred with the help of observations. In the following we decided to assume the source distribution derived by \cite{Fu2008} using data of the Canada-France-Hawaii Telescope Legacy Survey (CFHTLS). The distribution takes the functional form

\begin{equation}
n(z)=A\frac{z^{a}+z^{ab}}{z^{b}+c}~,
\end{equation}
with parameter values $a=0.612$, $b=8.125$, and $c=0.62$. The normalization constant $A$ is given by

\begin{equation}
 A^{-1}=\int_{0}^{+\infty}\frac{z^{a}+z^{ab}}{z^{b}+c}dz~.
\end{equation}
We checked our results against variations of the source redshift distribution, verifying that adopting instead the distribution of \citet*{Brainerd1996} (see also \citealt{Efstathiou1991,Smail1995}) or \cite{Benjamin2007} changed very little the subsequent cosmic shear results.

In Figure~\ref{fig:WLps} we show the effective convergence power spectra for the various oscillating dark energy models explored in this work, divided by the same quantity estimated in the framework of the concordance cosmology. We can observe that models 4 (yellow dot-dashed curve) and 6 (cyan dot-dotted curve) have negligible differences with respect to the $\Lambda$CDM model, while for the other models deviations can reach up to $\sim 10-15\%$ at very small angular scales. The fact that the largest deviations are visible at small scales highlights how the matter power spectrum is indeed affected by oscillating quintessence at those scales if one considers $z>0$.

Let us now make the case for a possible detection of oscillating dark energy more specific, by considering the S/N ratio for such a detection at a fixed multipole. This can be written as

\begin{equation}
\frac{S}{N}(\ell)=\left[\frac{P^\mathrm{DE}_\kappa(\ell)-P^{\Lambda\mathrm{CDM}}_\kappa(\ell)}{\Delta P^{\Lambda\mathrm{CDM}}_\kappa(\ell)}\right]^2~,
\end{equation}
where $\Delta P^{\Lambda\mathrm{CDM}}_\kappa(\ell)$ is the Gaussian statistical error on the power spectrum in the framework of the concordance cosmology. Following \cite{Kaiser1992,Kaiser1998,Seljak1998,Huterer2002}, the latter can be evaluated as

\begin{figure}
\includegraphics[angle=-90,width=0.45\textwidth]{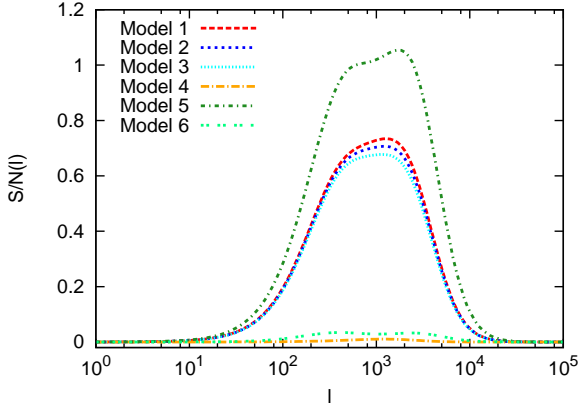}
\caption{The S/N ratio for discriminating between the concordance cosmology and each one of the six dynamical dark energy models considered in this work, as a function of multipole. Line styles and colours for the different models are as in Figure~\ref{fig:wz}.}
\label{fig:SN}
\end{figure}

\begin{figure*}
\includegraphics[angle=-90,width=0.33\textwidth]{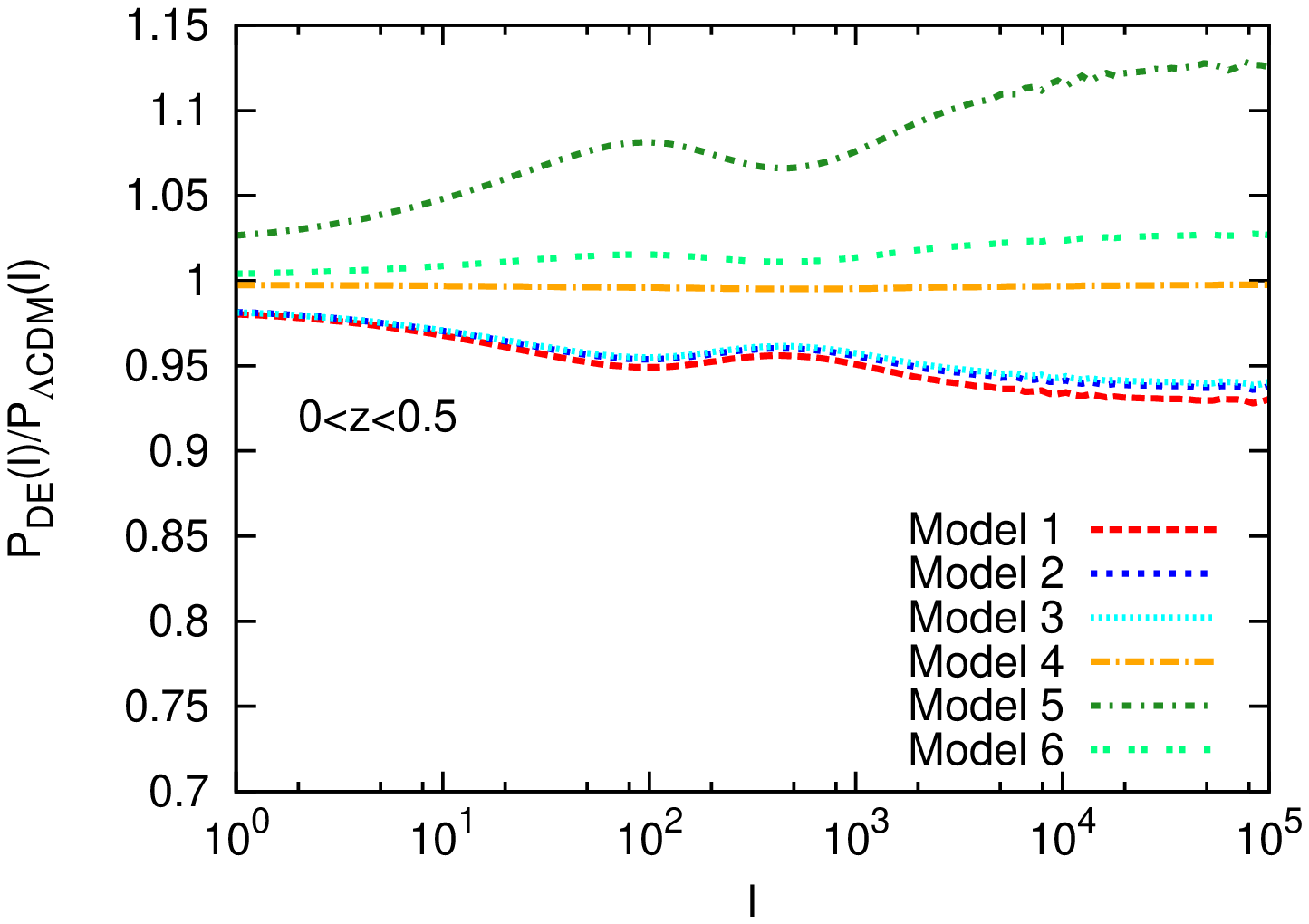}
\includegraphics[angle=-90,width=0.33\textwidth]{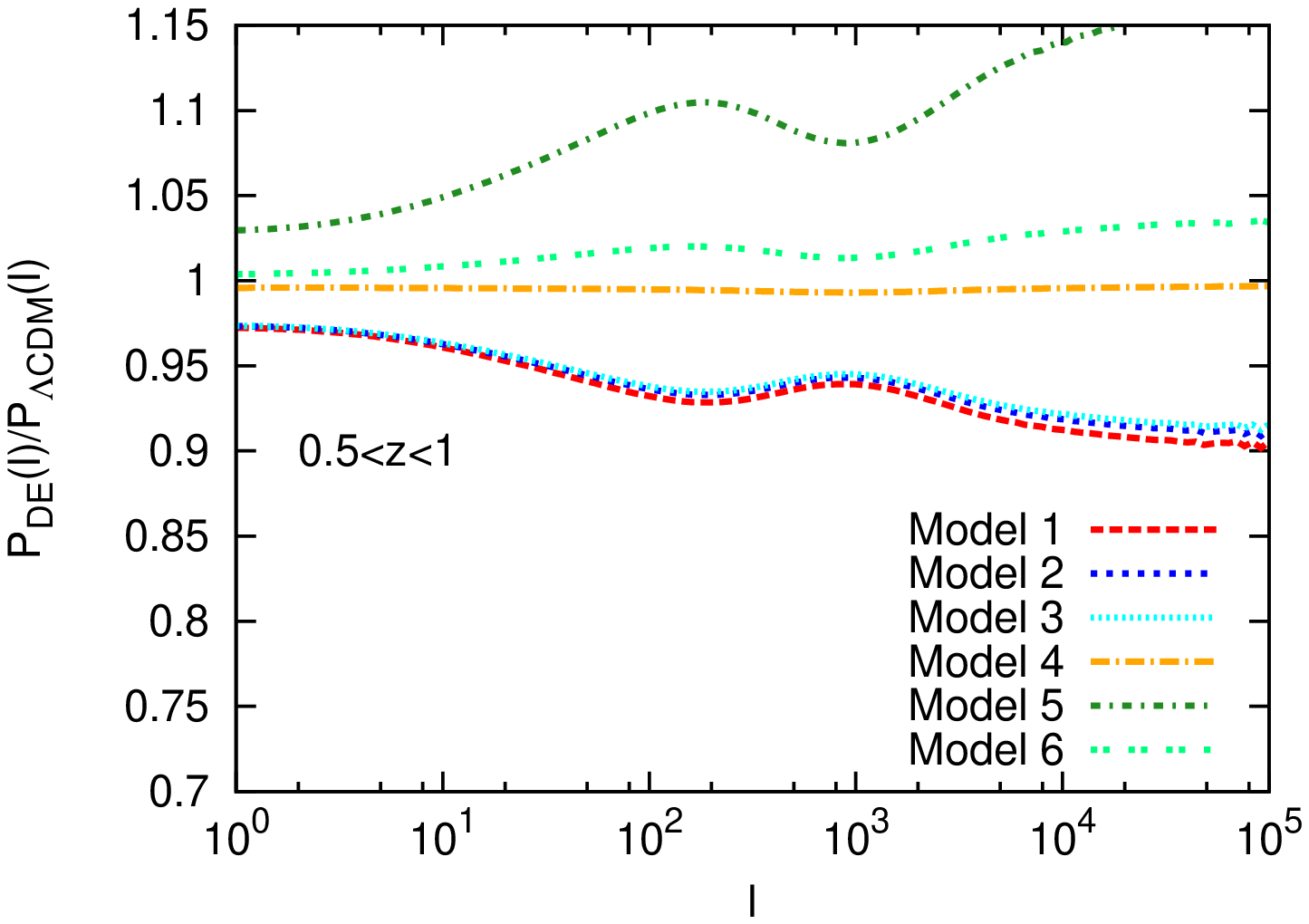}
\includegraphics[angle=-90,width=0.33\textwidth]{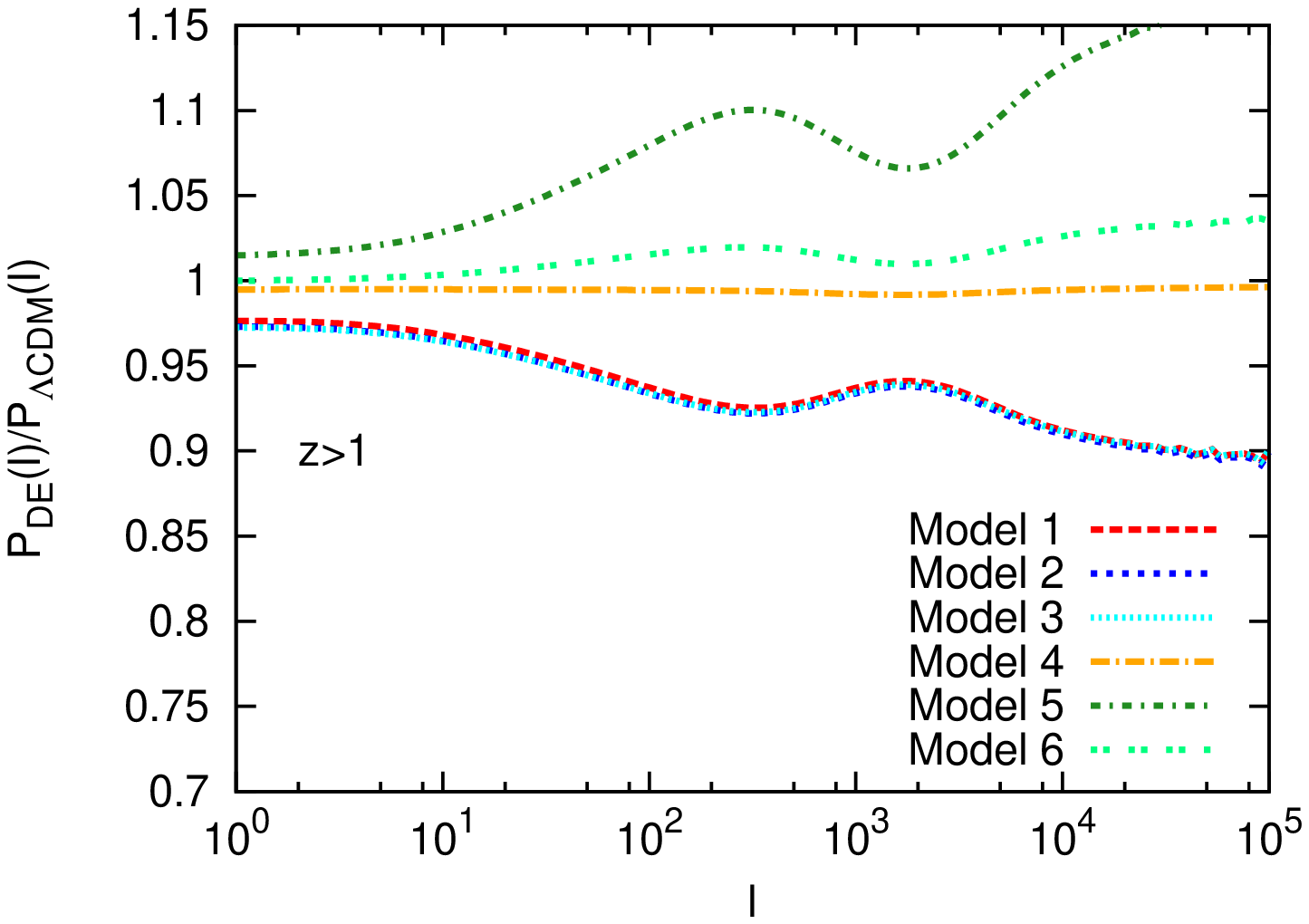}
\caption{The weak lensing power spectra for the six oscillating dark energy cosmologies computed by restricting to a specific source redshift bin. We show ratios with respect the $\Lambda$CDM expectation. Each panel refers to a particular redshift bin, as labeled. Line styles and colours for the different models are as in Figure~\ref{fig:wz}.}
\label{fig:tomography}
\end{figure*}

\begin{equation}
\Delta P^{\Lambda\mathrm{CDM}}_\kappa(\ell)=\sqrt{\frac{2}{(2\ell+1)\Delta\ell f_{\mathrm{sky}}}}\left[P^{\Lambda\mathrm{CDM}}_\kappa(\ell)+\frac{\gamma^{2}}{\bar{n}_\mathrm{g}}\right]~,
\end{equation}
where $\bar{n}_\mathrm{g}$ is the average surface number density of observed galaxies, $f_{\mathrm{sky}}$ is the fraction of sky area surveyed, and $\gamma$ represents the \emph{rms} intrinsic shape noise for the average galaxy. By assuming the specifications of a future typical weak lensing survey we set $\bar{n}=40$~arcmin$^{-2}$, $f_{\mathrm{sky}}=1/2$ and $\gamma=0.22$ \citep[see][]{Zhang2009}. To specify $\Delta\ell$, the binwidth over which the spectrum is averaged, we follow \cite{Takada2007} and \cite{Takada2009} and adopt the value $\Delta\ell=1$. A survey like \emph{Euclid} is expected to have similar performances \citep{Laureijs2011}.

We stress that the resulting S/N ratio values should be deemed accurate only for multipoles up to $\ell\sim 2000-3000$. For angular scales smaller than that non-Gaussian errors due to the non-linear growth of structures, which have not been taken into account in our analysis, kick in, and baryonic physics cannot be neglected anymore as well. Given this, in Figure~\ref{fig:SN} we show the S/N ratio for the effective convergence power spectrum. As also found by \cite{Fedeli2010} when studying the impact of primordial non-Gaussianity on the weak lensing power spectrum, we observe that at intermediate scales, $100\lesssim\ell\lesssim 1000$, S/N $\sim 0.5-1$ for four out of the six models we studied, while for very low or very high multipoles the S/N ratio tends to vanish. This means that for these models it would be sufficient to sum the S/N ratio over a relatively limited number of intermediate multipoles in order to have a significant detection of oscillating dark energy over the concordance cosmology. In models 4 and 6 however, the S/N ratio is at most at the level of $\sim 10^{-3}$, hence these models would be very difficult to distinguish from the $\Lambda$CDM cosmology by using a cosmic shear survey.

A maneuver usually capable of increasing the discriminatory power of cosmic shear consists in subdividing the source redshift distribution in a certain number of bins (usually $5$ is the maximum number that gives appreciable improvement, see \citealt{Sun2009}), computing the effective convergence power spectrum given by each bin, and then combining the various spectra together. This expedient is dubbed \emph{lensing tomography} \citep{Hu1999,Takada2004}. Specifically, the cross spectrum between the two source redshift bins $i$ and $j$ is just a straightforward generalization of the power spectrum defined above in Eq.~(\ref{eqn:wlps}),

\begin{equation}
P_\kappa^{ij}(\ell)=\frac{9H^{4}_{0}\Omega_{\mathrm{m},0}}{4c^{4}}
\int_{0}^{\chi_{\mathrm{H}}}P\left(\frac{\ell}{f_{\mathrm{K}}(\chi)},\chi\right)\frac{W_{i}(\chi)W_{j}(\chi)}{a^{2}(\chi)}d\chi~,
\end{equation}
where the integration kernels are now defined as

\begin{equation}
W_{i}(\chi)=\int_{\chi}^{\chi_{\mathrm{H}}}n_{i}(\chi^{\prime})\frac{f_{\mathrm{K}}(\chi-\chi^{\prime})}{f_{\mathrm{K}}(\chi^{\prime})}d\chi^{\prime}~.
\end{equation}
While previously the redshift distribution was normalized to unity over the complete redshift range, now we must normalize to unity the redshift distribution in each redshift bin, so that

\begin{equation}
\int_{0}^{\chi_{\mathrm{H}}} n_{i}(\chi)d\chi=1~.
\end{equation}

We did not attempt here a full tomographic analysis, since the resulting gain in discriminatory power is likely not enough to distinguish models 4 and 6, while the remaining models should be relatively easy to  distinguish by simply using the weak lensing power spectrum alone. We did however compute the power spectra resulting from three different source redshift bins, in order to verify whether one of them would give a markedly stronger signal than the others in order to focus observational efforts on that redshift range. The bins adopted are $[0,0.5]$, $[0.5,1]$, and $[1,+\infty]$. In Figure~\ref{fig:tomography} we show the ratio of the power spectra restricted to a specific source redshift bin computed for each of the six oscillating quintessence models in this work to the corresponding quantity evaluated in the framework of the $\Lambda$CDM cosmology. As can be seen, the qualitative behaviour is the same as for the full effective convergence power spectrum, although some quantitative differences exist. Specifically, the impact of oscillating dark energy is somewhat larger for the high redshift bins, having $z > 0.5$. However it is a relatively small effect, changing the deviations with respect to the $\Lambda$CDM cosmology of a few percent at most.

\section{Conclusions}\label{sect:conclusions}

In this work we considered structure growth in six different dark energy models characterized by an oscillating equation of state parameter $w(z)$. While many authors studied the expansion history of the Universe implied by these models, here we performed one step further by investigating the consequences of such models on the formation of non-linear structures. The main idea was to explore cosmological probes potentially capable of distinguishing one or more of these models from the concordance $\Lambda$CDM cosmology. To that aim, we studied several observables, ranging from the ISW effect to the cluster mass function and the cosmic shear power spectrum. Our main conclusions can be summarized as follows.

\begin{itemize}
\item No cosmologically relevant quantity shows oscillations as a function of redshift for any of the models considered in this work. This is a consequence of the fact that observables are given by integrals over $w(z)$, so that any feature in the latter function is efficiently smoothed out. There are very slight hints of a wiggle only in the ISW effect, in the age of the Universe, and in the deceleration parameter, at a level that is however likely impossible to detect.
\item We estimated the redshift drift, that is the variation of the cosmological redshift of a source due to the expansion of the Universe in the various models considered here. The impact of oscillating dark energy can reach up to $\sim 10\%$ at $z\sim 1$ for the most extreme cosmologies, and stays at the level of a few percent all the way up to $z\sim 10$. At these high redshifts the impact of peculiar motions is highly negligible, thus allowing to probe the expansion history with very high accuracy. 
\item The critical linear density contrast for spherical collapse $\delta_\mathrm{c}(z)$ and the virial overdensity $\Delta_\mathrm{v}(z)$ are always quite similar to the corresponding quantities evaluated within the fiducial cosmology, with deviations being at the level of a few percent for $z\lesssim 2-3$.
\item As naively expected, the growth factor of models having a larger amount of dark energy at early times is slightly larger than for the $\Lambda$CDM cosmology, and vice versa. Differences are at the level of a few percent also in this case and perfectly consistent with recent measurements presented by \cite{Blake2011a} for the redshift space distortions of the non-linear power spectrum.
\item The impact of oscillating quintessence on the mass function increases with both mass and redshift, reaching $\sim 30\%$ or more for $M > 10^{13} M_\odot/h$ at $z=2$. Interestingly, models with a high frequency of oscillations show an increment in the abundance of cosmic structures, while models with a more regular evolution of $w(z)$ show a decrement. This is because in models with high frequency of oscillations the amount of dark energy is higher than for the cosmological constant case and structures need to growth faster to compensate it (see Figures~\ref{fig:omega}, \ref{fig:gf} and \ref{fig:spc}). The resulting effect on the redshift distribution of cluster catalogues depend heavily on the selection function, however it is at a level likely to be detectable with future wide cluster surveys.
\item Cosmic shear is affected at the level of $\sim 10-15\%$ at intermediate/small angular scales. Given the sky coverage, sensitivity, and PSF stability of future wide field weak lensing surveys, at least some oscillating dark energy models will be discriminated form the concordance $\Lambda$CDM cosmology by using the power spectrum of effective convergence alone.
\end{itemize}

It is also worth to study what happens to our analysis if the dark energy equation of state is still oscillating, but is decreasing when the redshift increases in the vicinity of $z=0$, opposite to what we assumed now. In order to achieve that we added a phase $\theta=\pi$. We limited our new analysis only to the first three models and we consider the evolution in time of the growth factor, of the overdensities $\delta_{\mathrm{c}}$ and $\Delta_{\mathrm{V}}$ and the mass function at the four redshift considered, namely $z=0, 0.5, 1, 2$. We observed that for all the quantities considered, the behaviour is qualitatively the same but the effect is much smaller than before. We can therefore conclude that for the oscillating dark energy models, a first oscillation in the equation of state with decreasing values of the equation-of-state parameter makes the effects of dark energy smaller than before. This is due to the fact that the amount of dark energy in this situation is smaller than before, therefore structures do not need to grow as fast as it was before in order to compensate for it.

It is interesting to compare our results with those of \cite{Mignone2008}. In that work the authors showed that even if the expansion rate of the Universe has a sudden transition (that would require an even stronger transition in the dark energy equation of state parameter $w(z)$) this would hardly show up in cosmological observables such as the luminosity distance of SNe Ia. This agrees with our findings, namely since actual observables are given by at least a double integral over the function $w(z)$, any feature of the latter is easily smoothed out. A partial exception to this is the deceleration parameter. This is the second derivative of the expansion factor, hence it does retain some of the oscillatory behaviour of the dark energy equation of state parameter.

We conclude by noting that, although it is virtually not possible to find traces of oscillatory behaviour in cosmological observables, oscillating dark energy models do induce some specific modifications in the number counts of massive clusters and the power spectrum of cosmic shear that will likely be detectable by future cosmological surveys. This paper hence gives an additional contribution to the study of observational signatures of a dynamically evolving dark energy component, which is a fundamental field of study in order to better understand the nature of quintessence itself.

\section*{Acknowledgements}
This work was supported by the Deutsche Forschungsgemeinschaft (DFG) under the grants BA 1369/5-1 and 1369/5-2 and through the Transregio-Sonderforschungsbereich TR 33, as well as by the DAAD and CRUI through their Vigoni programme. We acknowledge financial contributions from contracts ASI-INAF I/023/05/0, ASI-INAF I/088/06/0, ASI I/016/07/0 COFIS, ASI Euclid-DUNE I/064/08/0, ASIUni Bologna-Astronomy Dept. Euclid-NIS I/039/10/0, and PRIN MIUR Dark energy and cosmology with large galaxy surveys. FP thanks also David Bacon, Robert Crittenden and Lado Samushia for useful discussions. We also thank the anonymous referee whose comments helped us to improve the presentation of our results.

\bibliographystyle{mn2e}
\bibliography{SmoothOsDE.bbl}

\appendix

\section{Code details and testing}\label{sect:code}

In this Appendix we explain in detail how we solved the Eqs.~(\ref{eqn:nleq}) and (\ref{eqn:leq}) for the non-linear and linear evolution of matter density fluctuations. Because all the terms involving the dark energy equation of state present oscillating behaviour, particular care has to be taken in order to achieve numerical convergence in the results. Since Eq. (\ref{eqn:leq}) is an ordinary second order linear differential equation, in order to solve it we need to provide the initial conditions $\delta_{\mathrm{i}}$ and $\delta^{\prime}_{\mathrm{i}}$. We assume that at early times the solution is a power law $\delta_{\mathrm{i}}=a^n$ (implying $\delta^{\prime}_{\mathrm{i}}=n\delta_{\mathrm{i}}/a$) and we insert this ansatz into the differential equation. By evaluating it at the equivalence scale factor $a_{\mathrm{eq}}$, we obtain a second order algebraic equation for $n$ which can be easily solved. Since at early times all the models are very well approximated by an EdS model, $n$ differs from unity only by a few percent, more so for models having more dark energy at early times. We parametrized the initial density contrast as a power law only in order to easily determine the velocity.

To solve Eq.~(\ref{eqn:nleq}) we have to provide as before two initial conditions, one for $\delta$ and one for $\delta'$. At early times, we can safely assume that the evolution of density perturbations is linear and the solution can be written in the same form as done before. As explained in the main text, we need to determine the initial overdensity making the perturbation diverge at a given time. To do so we fix the time (scale factor) corresponding to the collapse and run a root-search algorithm. At each step $\delta_{\mathrm{i}}$ and $\delta^{\prime}_{\mathrm{i}}$ are automatically updated. The values corresponding to the divergent result of the non-linear evolution are then used as initial conditions for the linear equation in order to determine the linear overdensity parameter $\delta_{\mathrm{c}}$. Despite the fact that, compared to \cite{Pace2010}, we do not have a fixed value for the initial velocity of the perturbations, we could reproduce the results presented there, implying that the role of the initial velocity is rather marginal. The advantages of doing so are twofold: on one side the new velocities are formally correct and can vary according to the cosmological model and on the other side the code is numerically much more stable. One instability problem is due to the fact that formally a divergent value needs to be infinite and this can not obviously be satisfied from the numerical point of view. We therefore assume that divergence occurs when $\delta\geqslant 10^{7}$. This makes $\delta_{\mathrm{c}}$ artificially increase with the redshift also for an EdS Universe (see Figure \ref{fig:spc}). We checked that having the initial velocity related to the initial overdensity makes this problem much less severe. We run the code to determine the value of $\delta_{\mathrm{c}}$ up to $z_{\mathrm{c}}=50$ and we saw that first of all the increase is very mild, and second the numerical value is higher than the analytical one by only $0.5\%$ at $z_{\mathrm{c}}=50$. For Eq.~(\ref{eqn:nleq}) the initial scale factor is $a_{\mathrm{i}}=5\times 10^{-5}$.

A crucial point for having numerical convergence of the results is to perform an accurate numerical integration for the equation determining the time evolution of the dark energy density. \\
To perform the integral in Eq.~\ref{eqn:g}, we made sure to have a numerical accuracy of at least $10^{-3}$ when compared with the analytical result. We also checked that different integration methods would give the same result as indeed was the case.

\label{lastpage}

\end{document}